\documentclass[letterpaper,11pt]{article}

\usepackage{amsmath,amssymb}
\usepackage{epsfig}
\usepackage{axodraw}

\begin{document}

\baselineskip=18pt

\newcommand{\eps}{\epsilon}
\newcommand{\pslash}{\!\not\! p}
\newcommand{\qslash}{\!\not\! q}
\newcommand{\ppslash}{\!\not\! p^{\,\prime}}
\newcommand{\m}{\widetilde m_u}
\newcommand{\M}{\widetilde M_u}
\newcommand{\Real}{\mathsf{Re}}
\newcommand{\e}[1]{\varepsilon_#1}


\thispagestyle{empty}
\vspace{20pt}
\font\cmss=cmss10 \font\cmsss=cmss10 at 7pt

\hfill
\vspace{20pt}

\begin{center}
{\Large \textbf
{The Minimal Composite Higgs Model \\ and Electroweak Precision Tests}}
\end{center}

\vspace{15pt}
\begin{center}
{\large Kaustubh Agashe$\, ^\text{$a$, $b$, $c$}$, Roberto Contino$\, ^a$} \vspace{20pt}

$^{a}$\textit{Department of Physics and Astronomy, Johns Hopkins University,
Baltimore, MD 21218, USA}

$^{b}$\textit{School of Natural Sciences, Institute for Advanced Study, Princeton, NJ 08540, USA}

$^{c}$\textit{Department of Physics, Syracuse University, Syracuse, NY 13244, USA}

\end{center}

\vspace{20pt}
\begin{center}
\textbf{Abstract}
\end{center}
\vspace{5pt} {\small \noindent
A complete analysis of the electroweak precision observables 
is performed within a recently proposed minimal composite Higgs model,
realized as a 5-dimensional warped compactification.
In particular, we compute
$Z\to b\bar b$ and the one-loop correction to the $\rho$ parameter.
We find that oblique data can be easily reproduced without a significant
amount of tuning in the parameters of the model, while $Z\to b\bar b$
imposes a stronger constraint. As a consequence of the latter, 
some of the new fermionic resonances must have mass around 4~TeV, 
which corresponds to an electroweak fine tuning of a few percent.
Other resonances, such as $Z'$, can be lighter
in sizeable portions of the parameter space.
We discuss in detail the origin of the $Z\to b\bar b$ constraint and we
suggest several possible avenues beyond the minimal model for weakening it.
}

\vfill\eject
\noindent


\section{Introduction: a composite Higgs from a fifth dimension}

Theories of the Higgs as a composite pseudo-Goldstone
boson (PGB) of a strongly interacting sector~\cite{GK} interpolate
between the two paradigms for electroweak symmetry breaking (EWSB):
technicolor~\cite{TC} and a fundamental scalar condensate.
Dimensional transmutation still elegantly solves the hierarchy problem, while
the potential discrepancies of technicolor (TC) with precision tests 
are avoided through a two-step symmetry breaking: 
at some scale $f_\pi$, the dynamical breaking of a global symmetry $G$ 
of the strong sector forms the Higgs doublet as a composite Goldstone boson; 
radiative loop corrections from $G$-violating interactions with an external sector
generate a potential for the Higgs, triggering EWSB at a scale $v\lesssim f_\pi$.
In the limit of a large separation between $v$ and $f_\pi$, $\eps= v/f_\pi \ll 1$, 
one recovers the Standard Model (SM) description, which means that all the corrections
to the electroweak precision observables are suppressed by powers of $\eps$.
In the opposite limit $\eps=1$, the phenomenology approaches that of a 
(non-minimal) TC theory, with
at least one light neutral boson in the spectrum, 
the physical Higgs, with an $O(1)$ quartic coupling.

The idea of the Higgs as a composite PGB
has recently found an interesting realization in a calculable extra-dimensional
scenario with a warped fifth dimension~\cite{Contino:2003ve,Agashe:2004rs}.
The holographic description of the 5D theory is that of a 
4D composite Higgs model, with two important new ingredients:
\textit{i) Conformality}: the strong sector is conformal at energies higher than
its mass gap, and it remains strongly coupled up to very high scales;
\textit{ii) Linear fermionic couplings}: elementary external
fermions couple linearly to the conformal sector (CFT) through composite fermionic operators~$O$.
The linear couplings between external fermions and strong sector realize
the partial compositeness scenario of Ref.~\cite{Kaplan:1991dc}.
Together with conformality, they offer an elegant solution to the 
flavour problem of the original Georgi-Kaplan composite models~\cite{GK}:
small differences in the anomalous dimensions of the operators $O$ can generate large 
hierarchies in the physical Yukawa 
couplings~\cite{Grossman:1999ra,Gherghetta:2000qt}, 
and suppress, at the same time, dangerous flavour changing neutral current (FCNC) effects 
from the strong dynamics through an extension of the GIM 
mechanism~\cite{Gherghetta:2000qt,Huber:2000ie,Agashe:2003zs,Agashe:2004cp}.
Moreover, due to conformality, linear couplings to the strong sector 
do not become highly irrelevant in the infrared (IR) 
even if the ultraviolet (UV) cut-off is 
Planckian (unlike the case of bilinear couplings in extended TC 
theories~\cite{Dimopoulos:1979es}), 
allowing us to extrapolate the theory up to the Planck scale.
Thus, FCNC effects from states at the cut-off are also negligible.

The most attractive feature of realizing the composite Higgs with an extra dimension
is calculability. In the limit of large 't Hooft coupling and large number $N$ of CFT 
colors, all coefficients and form factors of the 4D effective chiral lagrangian
of the $G$-symmetry breaking can be computed by resorting to the 5D picture. 
This in turn allows the computation of various physical observables, like for example 
the Higgs potential, in a $1/N$ expansion.
From the 5D viewpoint, calculability is guaranteed by a weakly coupled
field theory description, where perturbation theory in the 5D couplings
corresponds to the $1/N$ expansion of the 4D theory.
The connection between 4D composite Higgs models and 5D theories is made even more
intriguing by the idea of gauge-Higgs unification~\cite{gaugeHiggs}, 
since the role of the Higgs can be played (though not necessarily) by the fifth 
component $A_5$ of a 5D gauge field living in the bulk.

The virtue of 5-dimensional warped models,
besides a successful theory of flavour, is their UV-completeness up to
the Planck scale. This allows for a full explanation of the big 
hierarchy~\cite{Randall:1999ee} and opens up the possibility of having precision 
gauge coupling unification without supersymmetry~\cite{Agashe:2005vg}.
The same IR physics of electroweak symmetry breaking, on the other hand, can be captured 
by 5D effective models with a flat extra dimension and large brane kinetic terms
(see~\cite{Scrucca:2003ra}).

A minimal composite Higgs model (MCHM)~\footnote{
Notice that another model in a different context~\cite{Dobrescu:1999gv} adopted
the same name and acronym.} from a warped extra dimension
was introduced in Ref.~\cite{Agashe:2004rs}.
It is based on an SO(5)/SO(4) symmetry, thus featuring
an approximate custodial symmetry, and it has been shown to be 
a fully realistic playground to test the idea of composite Higgs.
The initial study of Ref.~\cite{Agashe:2004rs} carried out a complete calculation
of the Higgs potential and of the Peskin--Takeuchi $S$ parameter~\cite{PT},
presenting only naive dimensional estimates for the other two
important electroweak observables: $Z\to b\bar b$ and $\Delta\rho$.
The aim of the present work is to continue and complete that program
by performing a full detailed analysis of all electroweak precision tests (EWPT).
The success of the model will be measured by its capability of reproducing all known 
experimental results with a natural choice of parameters.

A model-independent analysis of the EWPT, performed without allowing
for correlations among different operators, showed that 
it might be difficult, in a generic extension of the Standard Model,
to reproduce all electroweak results and, at the same time, account for
a light Higgs without some degree of tuning~\cite{Barbieri:1999tm,Barbieri:2000gf}. 
How serious is this ``LEP paradox'' can be established by considering
specific models, like the MCHM.
The latter, like a generic composite Higgs theory, will eventually
pass all precision tests for 
$f_\pi$ sufficiently larger than the electroweak scale $v$.
Since one naturally expects $v\simeq f_\pi$,
the level of fine-tuning in the MCHM can be measured by how small
$\eps=v/f_\pi$ is required to be in order to satisfy the EWPT.
Values of $\eps\simeq 0.4-0.3$ naively suggest a $\sim 10\%$ cancellation among different
contributions in the Higgs potential, and were considered acceptable by 
Ref.~\cite{Agashe:2004rs}.

The same philosophy has been recently adopted 
by the authors of Ref.~\cite{Katz:2005au}.~\footnote{For another recent
proposal for a PGB Higgs with a similar amount of tuning, see 
Ref.~\cite{Chacko:2005pe}.}
Their ``intermediate'' Higgs 
can be thought of as 
the low-energy effective description of
a particular 4D composite Higgs theory
in which the first fermionic resonance of the strong sector is assumed
to be weakly coupled and lighter than the vector bound states.
Including this first fermionic resonance in the effective lagrangian,
and assuming a specific structure of interactions 
of the 
external quarks with the strong sector, 
the leading top quark contribution to the Higgs potential turns out 
to be calculable~\footnote{the gauge contribution to the Higgs potential is 
still UV-sensitive.} thanks to a collective breaking 
mechanism~\cite{Georgi:1975tz,Arkani-Hamed:2001nc}.
A different approach to the fine-tuning problem is proposed
by Little Higgs (LH) theories~\cite{Arkani-Hamed:2001nc,LH}, which try 
to generate naturally a large hierarchy between $v$ and $f_\pi$. 
The collective breaking is extended to the gauge sector as well,
and it is realized such as to suppress the size of the Higgs mass
term while still obtaining an $O(1)$ quartic coupling.
This makes $\eps$ naturally small.  However, this fact alone does not guarantee 
full success with EWPT. Indeed, although corrections
to the precision observables from the strong sector are now under control,
the additional weakly coupled light states needed to cutoff the quadratic divergence 
in the Higgs mass term generically reintroduce
large contributions to 
the precision observables~\cite{Csaki:2002qg, Hewett:2002px, recentLH}.
As proposed in~\cite{Cheng:2003ju}, 
a possible way to forbid these tree-level dangerous effects is by
introducing into the theory a discrete symmetry called T parity.

The same LH mechanism to generate a small $\eps$ -- namely: collective breaking 
\textit{plus} a differentiation between Higgs mass term and quartic coupling -- can also be
implemented in a 5D warped realization of the PGB, see~\cite{Thaler:2005en}.
The clear advantage in this way is that of having a UV completion up to the Planck 
scale.~\footnote{See Ref.~\cite{Katz:2003sn,Batra:2004ah} 
for different UV completions
of a Little Higgs theory.}
Indeed, when presented as effective descriptions valid up to a cutoff scale
$\Lambda\sim 5-10$ TeV, Little (and also Intermediate) Higgs models cannot
address important phenomenological issues like flavour, gauge
coupling unification or the big hierarchy.
Moreover, they are only \textit{technically} natural, since a specific
(and arbitrary) choice of interactions is usually needed to ensure the collective breaking.
In this respect, 5D composite Higgs models (as well as 5D LH constructions and other
schemes of EWSB with a UV completion) are more ambitious, hence more constrained,
as they aim to a complete explanation of the weak scale.
This means, in particular, that the interactions between external
fermions and strong sector will have the most general form allowed by gauge invariance.

Motivated by the above considerations, we present here a complete analysis
of the electroweak precision observables in the MCHM, hoping that what
is learned for this specific calculable model can 
be useful to better understand the more general class of composite Higgs theories.
After a brief review of the minimal model, we start by classifying all the 3-point
form factors needed to extract $Z\to b\bar b$ and $\Delta\rho$ (Section~\ref{sec:3pointff}).
The details of how to compute the relevant form factors using the holographic
technique of Ref.~\cite{Agashe:2004rs} can be found in Appendix.
Sections~\ref{sec:analysis} and~\ref{sec:closerlook} present the full analysis of EWPT,
and the spectrum of new particles is computed in Section~\ref{sec:spectrum}.
The consequences of our work are critically analyzed in the Conclusions.

\section{Computing $Z\to b \bar b$ and $\Delta\rho$ in the MCHM}
\label{sec:3pointff}

The minimal model introduced in Ref.~\cite{Agashe:2004rs}
is defined on the 5D spacetime metric~\cite{Randall:1999ee}
\begin{equation}
  ds^2 = \frac{1}{(kz)^2} \left(\eta_{\mu\nu}\, dx^\mu dx^\nu - dz^2\right)
  \equiv g_{MN}\, dx^M dx^N\, ,
\label{eq:metric}
\end{equation}
where the fifth dimension has two boundaries at $z=L_0 \equiv 1/k
\sim 1/M_{\rm Pl}$ (UV brane) and $z = L_1 \sim 1/{\rm TeV}$ (IR brane). 
Here and in the following we adopt the same notation 
as in Ref.~\cite{Agashe:2004rs}.
A gauge symmetry SU(3)$_c \times$SO(5)$\times$U(1)$_{B-L}$ 
of the 5D bulk is reduced, by boundary conditions,
to SU(3)$_c \times$SO(4)$\times$U(1)$_{B-L}$ on the IR brane 
(with SO(4)$\sim$SU(2)$_L \times$SU(2)$_R$),
and to SU(3)$_c \times$SU(2)$_L\times$U(1)$_Y$ on the UV brane.
In the unitary gauge $\partial_z (A_5/z)=0$, $A_5$ is non-vanishing
only in its SO(5)/SO(4) components, which describe 4D (physical) 
fluctuations with a fixed profile along the fifth dimension: 
$A_5^{\hat a}(x,z)= \zeta(z) h^{\hat a}(x)$, $\zeta(z)= 
z\sqrt{2/(L_1^2-L_0^2)}$~\cite{Contino:2003ve}.
The 4D scalar field $h(x)$ transforms as a {\bf 4} of SO(4)
and is identified with the Higgs field.
A potential for $h(x)$ is forbidden at tree level by locality
and the bulk gauge symmetry, but it is generated at the
radiative level through non-local finite effects.
This is the Hosotani mechanism for symmetry breaking~\cite{hosotani}.

Each SM quark generation is identified with the zero modes
of three 5D bulk Dirac spinor $\xi_i$
that transform as ${\bf 4_{1/3}}$ (spinorial) 
representations of SO(5)$\times$U(1)$_{B-L}$:
\begin{equation} \label{fstates}
\begin{split}
\xi_q &= \begin{bmatrix}
  q_L (++) & q_R (--)\\[0.1cm] Q_L(--) & Q_R (++) \end{bmatrix}\, , \quad
\xi_u = \begin{bmatrix}
  q_L^u (+-)  & q_{R}^{u}(-+) \\[0.1cm]
  Q_L^u = \begin{bmatrix} u^{c\, \prime}_L (-+) \\ d^{c\, \prime}_L (++)\end{bmatrix}
 &Q^u_R =\begin{bmatrix} u_R (+-) \\ d^{\,\prime}_R (--)\end{bmatrix}
        \end{bmatrix}\, , \\
\xi_d &= \begin{bmatrix}
  q_L^d (+-) & q_{R}^{d}(-+) \\[0.1cm]
  Q_L^d=\begin{bmatrix} u^{c\,\prime \prime}_L(++) \\ d^{c\, \prime \prime}_L (-+)\end{bmatrix}
 &Q^d_R =\begin{bmatrix} u^\prime_R (--) \\ d_R (+-)\end{bmatrix}
        \end{bmatrix}\, .
\end{split}
\end{equation}
Chiralities under the 4D Lorentz group have been denoted with $L,R$, while
small $q$'s (capital $Q$'s) denote doublets under SU(2)$_L$ (SU(2)$_R$).
The $\xi_i$ mix with an extra field $\widetilde Q_R$ localized on the
IR brane through the mass terms $[ \bar Q_L^u+ \bar Q_L^d]\widetilde Q_R$.
They also mix with each other through the most general 
SO(4)-invariant set of mass terms on the IR-brane
\begin{equation}
\label{massmixing}
[ \widetilde M_u
\bar Q_L^u +  \widetilde M_d
 \bar Q_L^d] Q_R +
\bar q_L[ \widetilde m_u
 q_R^u + \widetilde m_d  q_R^d] + \text{h.c.}
\end{equation}
Leptons are realized in a similar way.

A particularly useful way to match the above 5D theory to a
4D low-energy effective theory is by following the so-called
holographic description (see~\cite{Contino:2003ve,Agashe:2004rs} and also
\cite{Barbieri:2003pr,Contino:2004vy}), as opposed to the more 
conventional Kaluza-Klein (KK) decomposition.
It consists in integrating out the bulk (plus the IR-brane) dynamics 
and writing an effective action on the UV brane.
Boundary values of 5D fields with Neumann (Dirichlet) boundary
conditions on the UV brane will act like 4D dynamical fields
(external non-dynamical sources) of the brane effective 
action~\cite{Contino:2004vy,Agashe:2004rs}.
In the particular case of the fermion fields $\xi_i$, 
adopting a left-handed (right-handed) source description for 
$\xi_q$ ($\xi_{u,d}$)~\cite{Contino:2004vy},
the effective brane degrees of freedom can be organized in 
three 4D (chiral) spinorial representations of SO(5) with 
$(B-L)=1/3$:
\begin{equation}
\Psi_{q} = \begin{bmatrix} q_L \\[0.2cm] Q_L \end{bmatrix}\, , \qquad
\Psi_{u} = \begin{bmatrix} q^u_R \\[0.2cm]
             \begin{pmatrix} u_R \\ d'_R \end{pmatrix} \end{bmatrix}\, , \qquad
\Psi_{d} = \begin{bmatrix} q^d_R \\[0.2cm]
             \begin{pmatrix} u'_R \\ d_R \end{pmatrix} \end{bmatrix}\, .
\label{fermio}
\end{equation}
The dynamical fields $q_L$, $u_R$, $d_R$ match with the
quarks of a SM generation, while the additional components
$Q_L$, $q^u_R$, $d'_R$, $q^d_R$, $u'_R$ are 
non-dynamical spurion fields.
The latter do not play any physical role, but are a useful
tool to express the effective action in an 
(SO(5)$\times$U(1)$_{B-L}$)-invariant fashion.
Similarly, the gauge content of the effective action will
consist of complete adjoint representations $A_\mu$, $B_\mu$ of 
SO(5)$\times$U(1)$_{B-L}$, where however only the
gauge fields of SU(2)$_L \times$U(1)$_Y$ are truly dynamical.
By integrating out fluctuations of the Higgs field around
a constant classical background $\Sigma$, the effective
Lagrangian on the UV brane, in momentum space and at the
quadratic level, has the following structure~\cite{Agashe:2004rs}:
\begin{equation}
\begin{split}
{\cal L}_{\rm eff}^{(2)}
 =& \frac{1}{2} \left(P_T\right)_{\mu\nu}\Big[ \Pi^{B}_0(p)\, B^{\mu} B^{\nu}
    +\Pi_0(p)\, {\rm Tr}\big[A^{\mu} A^{\nu}\big]
    +\Pi_1(p)\, \Sigma A^{\mu} A^{\nu} \Sigma^T\Big] \\
  &+\sum_{r=q,u,d}\bar \Psi_r \pslash \Big[\Pi^{r}_0(p)
 +\Pi_{1}^r(p)\, \Gamma^i\Sigma_i \Big]\Psi_{r}
 +\sum_{r=u,d}\bar \Psi_q \big[M_0^r(p)+M_1^r(p)\, \Gamma^i\Sigma_i\big] \Psi_{r}\, .
\label{efflag}
\end{split}
\end{equation}
Here $(P_T)_{\mu\nu}=\eta_{\mu\nu}-p_\mu p_\nu/p^2$ and
$\Gamma^i$, $i=1,\dots 5$, are the gamma matrices for SO(5).
A possible mixing term between $\Psi_u$ and $\Psi_d$ in eq.~(\ref{efflag}) 
has been neglected since it does not play any role in our calculations.
Also, we have not included possible UV-brane kinetic and
gauge-fixing terms, i.e. terms not induced by the bulk dynamics. 
They can be included in a straightforward way.
The Goldstone field $\Sigma$ is parametrized by the fluctuations
along the (broken) SO(5)/SO(4) generators $T^{\hat a}$, $\hat a=1,2,3,4$:
\begin{equation}
\Sigma = \Sigma_0 e^{\Pi/f_\pi} \ ,
 \qquad \Sigma_0 =  (0,0,0,0,1) \ ,
 \qquad \Pi= -i T^{\hat a} h^{\hat a} \sqrt{2} \, ,
\end{equation}
so that
\begin{align} 
\Sigma &= \frac{\sin h/f_\pi}{h} \left( h^1,h^2,h^3,h^4, h \cot h/f_\pi \right) \, ,
 \qquad  h = \sqrt{ (h^{\hat a})^2 } \, , \label{eq:Sigma} \\[0.2cm]
\Gamma^i\Sigma_i &= 
  \begin{pmatrix}  
    \mathbf{1}\, \cos h/f_\pi & \hat\sigma\, \sin h/f_\pi \\
    \hat\sigma^\dagger\, \sin h/f_\pi & -\mathbf{1}\, \cos h/f_\pi
  \end{pmatrix} \, , \qquad
   \hat\sigma \equiv \sigma^{\hat a}\, h^{\hat a}/h\, , \quad
   \sigma^{\hat a} = \{ \vec \sigma,-i\mathbf{1} \}\, .
\end{align}

The structure and the symmetries of the effective action (\ref{efflag}) are exactly 
those one would have obtained by integrating out a 4D strongly interacting sector,
coupled to the external fields $\Psi_{q,u,d}$, $A_\mu$ and $B_\mu$,
in which an SO(5) flavor symmetry is spontaneously
broken down to SO(4) by a composite field $\Sigma$.
In fact, we obtained the effective action (\ref{efflag}) by integrating out the bulk dynamics,
and this suggests that the bulk is indeed acting like a 4-dimensional strongly
interacting sector.
The reduction of SO(5) down to SO(4) on the IR brane
corresponds to the spontaneous breaking of SO(5) in the strong sector, and
the non-vanishing components of $A_5$ correspond to the 4D Goldstone modes.
Most importantly, the analogy also implies that all the isometries and gauge
symmetries of the 5D bulk must be reflected in global symmetries of the
4D strong sector. Since in our case the former is a slice of AdS$_5$,
the latter will be conformal at high energies.
This holographic correspondence between the 5D theory and a 4D 
theory with a strongly interacting sector proves to be extremely useful
to have a quick understanding of the 5D physics, since it is based only on
symmetry arguments.
Moreover, the AdS/CFT correspondence~\cite{Maldacena:1997re} seems to suggest 
that the analogy can be promoted to an exact duality in a string framework.

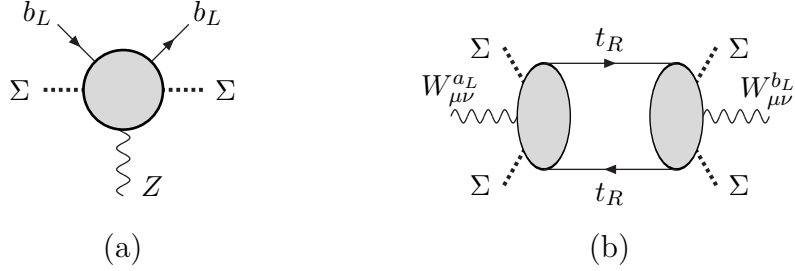
\begin{figure}[t]
\centering
   \begin{picture}(120,80)
     \Photon(60,50)(60,10){2.5}{5}
     \ArrowLine(70.6,60.6)(84.7,74.7) \ArrowLine(35.25,74.7)(49.4,60.6)
     \SetWidth{1.5} \DashLine(60,50)(30,50){1.5} \DashLine(60,50)(90,50){1.5}
     \SetWidth{1} \GCirc(60,50){15}{0.85} \SetWidth{0.5}
     \Text(67,10)[lb]{$Z$}
     \Text(33.25,74.7)[rb]{$b_L$} \Text(86.7,74.7)[lb]{$b_L$} 
     \Text(25,50)[r]{$\Sigma$} \Text(95,50)[l]{$\Sigma$}
     \Text(60,-10)[c]{\large (a)}
   \end{picture} \hspace{2cm}
   \begin{picture}(120,80)
     \SetWidth{1.5} \DashLine(35,40)(20,14){1.5} \DashLine(35,40)(20,66){1.5}
                    \DashLine(85,40)(100,66){1.5} \DashLine(85,40)(100,14){1.5}
     \SetWidth{1} \GOval(35,40)(20,10)(0){0.85} \GOval(85,40)(20,10)(0){0.85} \SetWidth{0.5}
     \ArrowLine(35,60)(85,60) \ArrowLine(85,20)(35,20)
     \Photon(0,40)(25,40){2.5}{4} \Photon(95,40)(120,40){2.5}{4}
     \Text(0,50)[c]{$W_{\mu\nu}^{a_L}$}  \Text(120,50)[c]{$W_{\mu\nu}^{b_L}$}
     \Text(15,14)[r]{$\Sigma$} \Text(15,66)[r]{$\Sigma$}
     \Text(105,66)[l]{$\Sigma$} \Text(105,14)[l]{$\Sigma$}
     \Text(60,65)[b]{$t_R$} \Text(60,15)[t]{$t_R$}
     \Text(60,-10)[c]{\large (b)}
   \end{picture} \vspace{0.7cm}
\caption{\it Diagrams in the 4D holographic theory that generate
the correction to $Z\to b_L \bar b_L$~(a), and $\Delta\rho$~(b). 
A grey blob represents the 4D CFT dynamics or, equivalently, the 5D bulk.} 
\label{fig:4Ddiagrams}
\end{figure}

These considerations show that our 5D model, through its holographic
description (\ref{efflag}), can be truly regarded and studied as a 4D composite Higgs model.
The great advantage of having a (weakly coupled) 5D description is that it
enables us to perform calculations.
The 2-point form factors $\Pi_i(p)$, $M_i(p)$ 
of eq.(\ref{efflag}) were computed in Ref.~\cite{Agashe:2004rs} 
in terms of 5D propagators, and this in turn 
allowed to compute the Higgs potential, the fermion masses and the  
Peskin-Takeuchi $S$ parameter.
We now want to extend the analysis of Ref.~\cite{Agashe:2004rs} by
computing $\Delta\rho$ (or equivalently the Peskin-Takeuchi
$T$ parameter) and the correction $\delta g_{L b}$
to the coupling of the left-handed bottom quark to the $Z$ boson.
Using the language of the 4D holographic picture, the leading contributions
to these observables come from the diagrams of Fig.\ref{fig:4Ddiagrams}.
In order to compute these diagrams we need to determine the 3-point form factors between
two fermions and a gauge field in the effective brane action.

The most general (SO(5)$\times$U(1)$_{B-L}$)-invariant effective Lagrangian 
that describes cubic interactions among two fermions in the spinorial representation
and a gauge field is,
in momentum space,~\footnote{Other possible form factor structures, like for example
$\bar\Psi_r \Gamma^{\mu ,\, r}_4(p,p') \left(A_\mu-\Sigma A_\mu\Sigma \right) \Psi_r$ or
$\bar\Psi_r \Gamma^{\mu ,\, r}_5(p,p') \{T^A ,\Sigma \} \Psi_r \; \Sigma \{A_\mu,T^A \} \Sigma^T $,
can be expressed in terms of those in eq.(\ref{efflag3}).}
\begin{equation} \label{efflag3}
\begin{split}
{\cal L}^{(3)}_{eff} = \sum_{r=q,u} \bigg\{
  &\bar\Psi_r \left( \Gamma^{\mu ,\, r}_0(p,p')A_\mu 
               +\frac{1}{2} \Gamma^{\mu ,\, r}_1(p,p') \{A_\mu,\Sigma\} \right) \Psi_r
  + \bar\Psi_r \Gamma^{\mu ,\, r}_2(p,p') T^A \Psi_r \; \Sigma \{A_\mu,T^A \} \Sigma^T \\
 &+ \bar\Psi_r \Gamma^{\mu ,\, r}_3(p,p')\, i[A_\mu,\Sigma] \Psi_r 
  + q_{(B-L)}\, B_\mu \,  \bar\Psi_r \left( \Gamma^{\mu ,\, r}_{0 B}(p,p') 
               + \Gamma^{\mu ,\, r}_{1 B}(p,p') \Sigma \right) \Psi_r \bigg\}\, .
\end{split}
\end{equation}
Here $q_{(B-L)}=1/3$, $p$ and $p'$ stand for the 4D momenta of the two fermions, 
and $\Sigma$ is treated like a classical constant background.
For simplicity, we have not included $\Psi_d$ in the sum over fermionic species, 
assuming that its effects in $\Delta\rho$ and $\delta g_{Lb}$ can be neglected.
This could be the case, for example, if the mixing of $\Psi_d$ with the other two bulk
multiplets is slightly suppressed.
Also, we have not written down mixing terms between $\Psi_q$ and $\Psi_u$, because they
are irrelevant to the computation of $\Delta\rho$ and $\delta g_{Lb}$.
The form factors $\Gamma^\mu$ satisfy the Hermiticity condition
\begin{equation} \label{hermiticity}
\gamma^0 \Gamma^{\mu\,\dagger}(p',p) \gamma^0 = \Gamma^\mu(p,p')\, ,
\end{equation}
and can be decomposed as linear combinations of the following complete 
set of Lorentz structures:
\begin{equation}
\Big\{ \gamma^\mu, \, i\epsilon^{\mu\alpha\beta\delta}p_\alpha p'_\beta \gamma^\delta ,\,
(p^\mu \pslash + p^{\prime \mu}\ppslash ) , \, (p^\mu \ppslash + p^{\prime \mu}\pslash ) , \,
i(p^\mu \pslash - p^{\prime \mu}\ppslash ) , \, i(p^\mu \ppslash - p^{\prime \mu}\pslash ) \Big\}\, .
\end{equation}
The form factors $\Gamma^\mu_{0,1}$, $\Gamma^\mu_{0 B,1 B}$ are related by the Ward identities
to the 2-point form factors $\Pi_{0,1}$ of eq.(\ref{efflag}):
\begin{equation} \label{WI}
(p-p')_\mu \Gamma^\mu_{i}(p,p') = (p-p')_\mu \Gamma^\mu_{i B}(p,p')
 = \pslash \, \Pi_{i}(p) - \ppslash \,\Pi_{i}(p')\, , \qquad i=0,1\, .
\end{equation}
Another useful form of the Ward identities is obtained in the limit $p=p'$:
\begin{equation}
\Gamma^\mu_{i}(p,p) = \Gamma^\mu_{i B}(p,p) =
\gamma^\mu \Pi_{i}(p) + 2\pslash \, p^\mu \frac{d}{dp^2} \Pi_{i}(p)\, , \qquad i=0,1\, .
\end{equation}

All the form factors of eq.(\ref{efflag3}) can be easily computed in the 5D picture 
by using the holographic procedure of Ref.~\cite{Agashe:2004rs}.
The details of the computation and the final expressions of the $\Gamma^\mu$'s
in terms of 5D propagators are given in Appendix.
Here we just notice that the term proportional to $\Gamma^\mu_3$ identically vanishes
if eq.(\ref{efflag3}) is evaluated upon only physical states (which means that 
$\Gamma^\mu_3$ will not appear in $\delta g_{Lb}$ or $\Delta\rho$), but it must be considered 
when extracting the other form factors with the holographic method used in Appendix.

Having classified all the possible 3-point form factors, we are ready to compute 
$\delta g_{Lb}$ and $\Delta\rho$ from the diagrams of Fig.\ref{fig:4Ddiagrams}.
Let us start with $\delta g_{Lb}$. We set the Higgs to the physical vacuum,
\begin{equation}
\langle \frac{h^{\hat a}}{h} \rangle =\delta^{\hat a 3}\, , \qquad
\langle \Sigma \rangle = \big(0,0,\eps,0,\sqrt{1-\eps^2}\big)  \, , \qquad
\eps\equiv \frac{v}{f_\pi} = \sin \frac{\langle h\rangle}{f_\pi}\, ,
\end{equation}
and consider the terms in eqs.(\ref{efflag}),(\ref{efflag3}) that involve the physical fields 
$A_\mu^{3_L}$, $Y_\mu \equiv A_\mu^{3_R} = 2 B_\mu$, $\Psi_q= (q_L ,0)$.
After rescaling the fields to canonically normalize their kinetic term, one has:
\begin{equation}
\begin{split}
{\cal L}_{eff}^{(3)} \supset \;
& \bar q_L \bigg[
    g A_\mu^{3_L} \frac{\sigma_3}{2} \;
      \frac{\Gamma^{\mu ,\, q}_0+ \Gamma^{\mu ,\, q}_1 \sqrt{1-\eps^2}}%
           {\Pi^q_0 + \Pi^q_1 \sqrt{1-\eps^2}}
  + g' Y_\mu \frac{1}{6} \;
      \frac{\Gamma^{\mu ,\, q}_{0 B}+ \Gamma^{\mu ,\, q}_{1 B} \sqrt{1-\eps^2}}%
           {\Pi^q_0 + \Pi^q_1 \sqrt{1-\eps^2}}
  \bigg] q_L \\
&+ \bar q_L \bigg[
   \left(g A_\mu^{3_L}-g' Y_\mu \right) \frac{\eps^2}{2} \frac{\sigma^3}{2} \,
      \frac{\Gamma^{\mu ,\, q}_2}%
           {\Pi^q_0 + \Pi^q_1 \sqrt{1-\eps^2}}
  \bigg] q_L \, .
\end{split}
\end{equation}
To extract the physical coupling of the bottom quark to the $Z$, we need to go on shell.
A good approximation is to neglect $m_b$ and $m_Z$, and set $p^2=p^{\prime\, 2}=(p-p')^2=0$ 
(the error is respectively of order ${\cal O}(m_b^2/m_\rho^2)$, ${\cal O}(m_Z^2/m_\rho^2)$,
with $m_\rho\sim 4\pi f_\pi/\sqrt{N}$).
From their explicit expressions in Appendix, one can show that on shell 
the form factors $\Gamma^{\mu}_i$ are proportional to $\gamma^\mu$:
\begin{equation}
\begin{split}
& \Gamma^{\mu}_{i} \big|_\text{on shell}=\Gamma^{\mu}_{i B} \big|_\text{on shell}= 
 \gamma^\mu\, \Pi_{i}(0)\, , \qquad i=0,1\, , \\
& \Gamma^{\mu}_2 \big|_\text{on shell} = \gamma^\mu\, \Gamma_{2}(0)\, ,
\end{split}
\end{equation}
where $\Gamma_{2}(p)$ is given in Appendix and $\Pi_{0,1}(p)$ can be found in 
Ref.~\cite{Agashe:2004rs}. Using the above relations, it is easy to derive the correction 
to the coupling of $b_L$ to $Z$ as compared to the SM value:
\begin{equation} \label{dgLb}
\delta g_{Lb} = -\frac{\eps^2}{4} \frac{\Gamma^{q}_{2}(0)}{\Pi^q_0(0)+\Pi^q_1(0) 
 \sqrt{1-\eps^2}}\, .
\end{equation}

We now turn to the computation of $\Delta\rho$, as defined by
\begin{equation}
\Delta\rho = \alpha T = \frac{4}{v^2} \left(\Pi_{11}(0)-\Pi_{33}(0) \right)\, ,
\end{equation}
in terms of the vacuum-polarization amplitudes
$\Pi^{\mu\nu}_{ij}(q^2)=\eta^{\mu\nu} \Pi_{ij}(q^2)+q^\mu q^\nu \hat \Pi_{ij}(q^2)$
for the SU(2)$_L$ gauge fields.
The leading effect that violates the custodial symmetry, and thus contributes
to $\Delta\rho$, comes from the diagram of Fig.\ref{fig:4Ddiagrams}(b). 
It corresponds to a contribution
to $\Pi_{33}(0)$ only, since both fermions in the loop are $t_R$.
To extract the effective vertex $A_{\mu}^{3_L} \bar t_R t_R$ we consider the terms in 
eqs.(\ref{efflag}),(\ref{efflag3}) that involve $A_\mu^{3_L}$ and $t_R$:
\begin{equation}
{\cal L}_{eff} \supset 
 \bar t_R \pslash \left[ \Pi_0^u - \Pi_1^u \sqrt{1-\eps^2}\right] t_R
 -\frac{\eps^2}{4}\, \bar t_R A_\mu^{3_L} \Gamma^{\mu ,\, u}_2 t_R\, .
\end{equation}
Since we are only interested in $\Pi_{33}(q^2=0)$, we can take the limit of zero 
external momentum in the diagram of Fig.\ref{fig:4Ddiagrams}(b). In this limit
the relevant 3-point form factor has the following structure:
\begin{equation} \label{G23}
\Gamma^{\mu ,\, u}_2(p,p) =
 \gamma^\mu \, \Gamma^u_{2}(p) + 2 \pslash \, p^\mu \Delta^u_{2}(p)\, ,
\end{equation}
where $\Delta^u_{2}(p)$ is given in Appendix.
Computing the loop amplitude we then find, in the Euclidean,
\begin{equation} \label{drho}
\Delta\rho = \alpha T = \frac{N_c}{v^2}\, \frac{\eps^4}{4}
 \int \!\! \frac{d^4p}{(2\pi)^4} \frac{1}{p^2}\,
 \frac{(\Gamma^u_{2}(ip))^2-2 p^2 \Delta^u_{2}(ip)\Gamma^u_{2}(ip) -2 p^4 (\Delta^u_{2}(ip))^2}%
      {\left( \Pi^u_0(ip)-\Pi^u_1(ip)\sqrt{1-\eps^2} \right)^2}\, ,
\end{equation}
where $N_c=3$ stands for the number of QCD colors.

\section{Complete analysis of EWPT}
\label{sec:analysis}

Universal electroweak corrections (from the SM and New Physics) to the precision observables
measured by LEP1 and SLD experiments~\cite{LEPEWWG} can be efficiently and fully summarized 
in terms of three parameters: $\e1 ,\e2, \e3$~\cite{eps123}.
A fourth parameter, $\e{b}$, can be added to describe non-universal effects in the bottom quark
sector~\cite{epsb}. 
The $\e{i}$ are related to $S$, $\Delta\rho$, $\delta g_{Lb}$ as follows:
\begin{subequations} \label{eq:pred}
\begin{align}
\e{1} &= \left( +5.60 - 0.86\, \ln\frac{m_H}{M_Z} \right)\cdot 10^{-3} + \Delta\rho \, , \\
\e{2} &= \left( -7.09 + 0.16\, \ln\frac{m_H}{M_Z} \right)\cdot 10^{-3} \, , \\
\e{3} &= \left( +5.25 + 0.54\, \ln\frac{m_H}{M_Z} \right)\cdot 10^{-3} + 
 \frac{\alpha}{4 s_W^2}\, S \, ,\\
\e{b} &= -6.43\cdot 10^{-3} - 2\, \delta g_{Lb} \, .
\end{align}
\end{subequations}
The first term in each of the above equations represents an approximation 
of the SM contribution (accurate for 
not too light Higgs masses, $m_H\gtrsim 50$ GeV), 
as computed using the code TopaZ0~\cite{TopaZ0} with 
$m^{pole}_t= 172.7\,$GeV.~\footnote{See Ref.~\cite{Barbieri:2004qk}. We thank Alessandro Strumia 
for providing us with numbers updated to $m_t^{pole}=172.7$ GeV.}
We have neglected the contribution to the $\e{i}$ that is encoded in the two additional 
parameters $W$,~$Y$ defined in Ref.~\cite{Barbieri:2004qk}, since 
in our model $W$,~$Y$ are 
suppressed by a factor $(g^2 f_\pi^2/m_\rho^2)\sim (g^2 N/16\pi^2)$ 
compared to $S$ and $T$.
This also implies that the LEP2 results do not pose strong constraints on the
parameters of our model, and can be neglected.

A fit to the $\e{i}$ using the LEP1 and SLD results gives~\cite{Ale}:
\begin{equation} \label{eq:fit}
\begin{split}
\e{1} =& (+4.9\pm 1.1)\cdot 10^{-3} \\
\e{2} =& (-9.1\pm 1.2)\cdot 10^{-3} \\
\e{3} =& (+4.8\pm 1.0)\cdot 10^{-3} \\
\e{b} =& (-5.2\pm 1.5)\cdot 10^{-3} 
\end{split}\qquad\quad
\rho = \begin{pmatrix}
1     & 0.59  & 0.83  & -0.28 \\
0.59  & 1     & 0.45  & -0.13 \\
0.83  & 0.45  & 1     & -0.16 \\
-0.28 & -0.13 & -0.16 & 1
\end{pmatrix}\, ,
\end{equation}
where $\rho$ is the correlation matrix.
An analogous fit performed by the LEP Electroweak Working Group leads to
similar results~\cite{LEPEWWG}.
By using eqs.(\ref{eq:pred}) and (\ref{eq:fit}) we can perform a detailed test of the electroweak
corrections in our model. 

We carried out a numerical analysis of the MCHM by scanning over the parameter space of 
the theory: for each point we extract $N$, the number of colors of the
CFT sector, by fixing the top quark mass to its experimental value; 
evaluate the Higgs effective potential, determining 
$\eps \equiv v/f_\pi = \sin\langle h\rangle/f_\pi$ and the Higgs mass $m_H$; 
compute $S$, $\Delta\rho$, $\delta g_{Lb}$ 
by using eqs.(\ref{dgLb}),(\ref{drho}) and the formulas 
given in Ref.~\cite{Agashe:2004rs}. Explicit expressions for the potential and the top quark
mass can also be found in Ref.~\cite{Agashe:2004rs}.
The 5D input parameters are the bulk SO(5) gauge coupling $g_5$, the
gauge kinetic terms on the UV and IR branes, $1/g^2_\text{UV}$ and
$1/g^2_\text{IR}$ (respectively for SU(2)$_L$ and SO(4)), 
and the top bulk and IR-brane masses, $c_q$, $c_u$, $\m$ and $\M$.
We varied $ -0.3 \leq c_u \leq 0.3$, $0.36 \leq c_q \leq 0.45$, 
and $0 \leq \m \leq 2$, $-4 \leq \M \leq -2$.
The UV coupling $g_{UV}$ is fixed by requiring that the low-energy
SU(2)$_L$ gauge coupling $g$ equals its experimental value, while 
the IR gauge kinetic term has been set to be of loop
order: $1/g^2_\text{IR}=\delta_{IR}/16\pi^2$, where we varied $1\leq \delta_{IR} \leq 3$.
The SO(5) bulk coupling $g_5$ defines $N$, $1/N \equiv g_5^2 k/16\pi^2$,
which is in turn fixed by the top mass, as we said above.
For the latter we adopted the new measurement 
$m^{pole}_t = (172.7 \pm 2.9)\,$GeV~\cite{Group:2005cc}, evolved
to $\mu =2\,$TeV, a typical scale at which the
bound states of the strong sector form, and converted to the
$\overline{\text{MS}}$ scheme: 
$m_t^{\overline{\text{MS}}}(2\,\text{TeV})= 150\,$GeV.~\footnote{Ref.~\cite{Agashe:2004rs}
used $m^{pole}_t = (178.0 \pm 4.3)\,$GeV and did not include the RG running, 
thus obtaining $m_t^{\overline{\text{MS}}}= 169\,$GeV.}

Having included the running of the top Yukawa coupling in $m_t$, for
consistency we need to consider the effects of the QCD dressing also in the 
Higgs potential. From the 4D holographic viewpoint, the leading effect is in the
renormalization of the couplings $\lambda_u$, $\lambda_q$ of the elementary 
$t_R$, $q_L$ to the CFT. 
We have estimated that this correction is negligible at energies above
$\mu \sim 2\,$TeV, since $\lambda_u$ and $\lambda_q$  flow rapidly 
to an attractive fixed point value.
Below $\mu$, the running of $\lambda_u$, $\lambda_q$ in the diagrams contributing to
the Higgs potential corresponds to that of the top Yukawa $y_t$.
It can be included, for example, by matching the potential at the scale
$\mu$ and evolving its coefficients down to low energies,
using the SM field content as an effective theory description.
Since the QCD corrections make the top Yukawa larger in the IR,
this leads to a slightly larger physical Higgs mass.
We have included this effect in our analysis by adding the following
correction to the Higgs mass squared:
\begin{equation}
\Delta m_H^2 = 2 v^2 \frac{2 N_c}{16\pi^2} 
 \int^{2\, \text{TeV}}_{m_t^{pole}} \! d\log\mu \,
 \left[ y_t^4(\mu) - y_t^4(2\, \text{TeV}) \right] = \left( 35\, \text{GeV}\right)^2\, .
\end{equation}

\begin{figure}
\begin{minipage}[t]{0.95\linewidth}
\epsfig{file=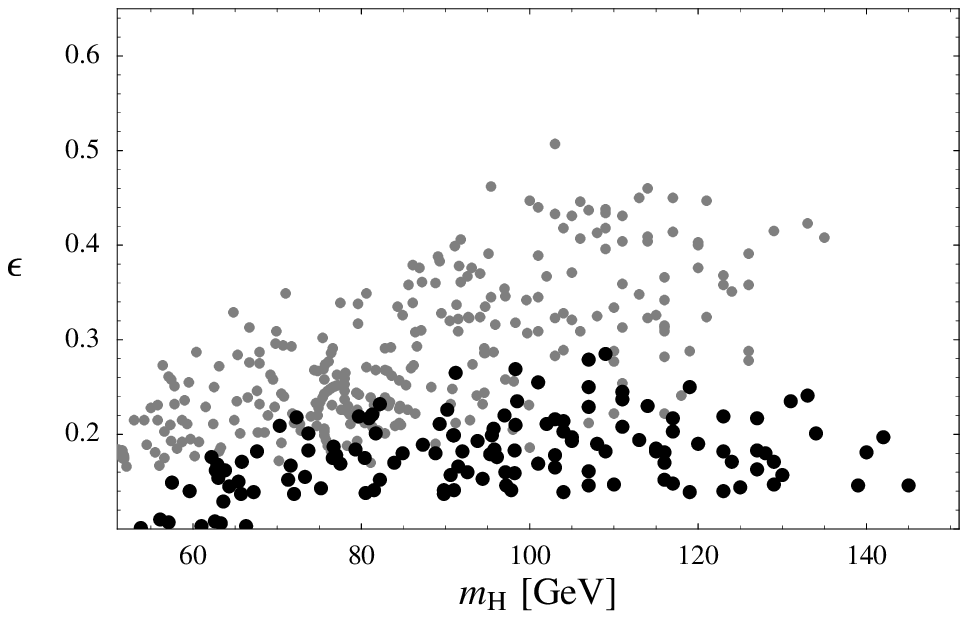,width=0.498\linewidth} \qquad
\epsfig{file=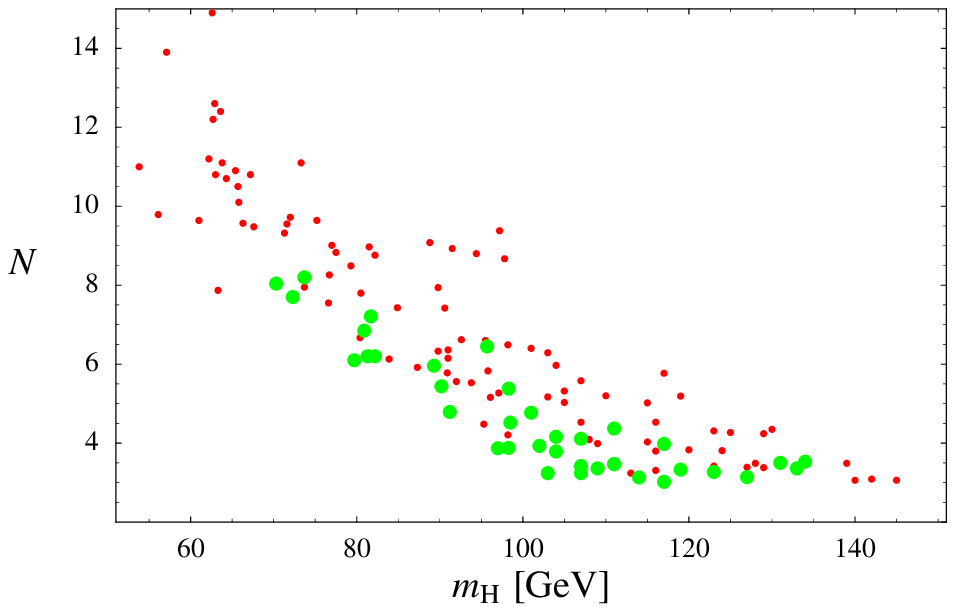,width=0.498\linewidth}
\end{minipage}
\caption{\it Scatter plots in the $(m_\text{Higgs},\eps)$ plane (left) and in the 
$(m_\text{Higgs},N)$ plane (right), obtained by scanning over the input parameters
of the MCHM. All points shown pass the $\chi^2$ test ($\Delta\chi^2<13.28$, $N\geq 3$), 
except for the grey points in the first plot, which satisfy the weaker constraint 
$|\delta g_{Lb}/g_{Lb}^{SM}| \leq~1\%$ (see text). In the second plot, 
green fat dots correspond to $0.2 < \eps < 0.3$, small red dots
to $\eps < 0.2$.}
\label{fig:full}
\end{figure}

\begin{figure}
\begin{minipage}[t]{0.95\linewidth}
\epsfig{file=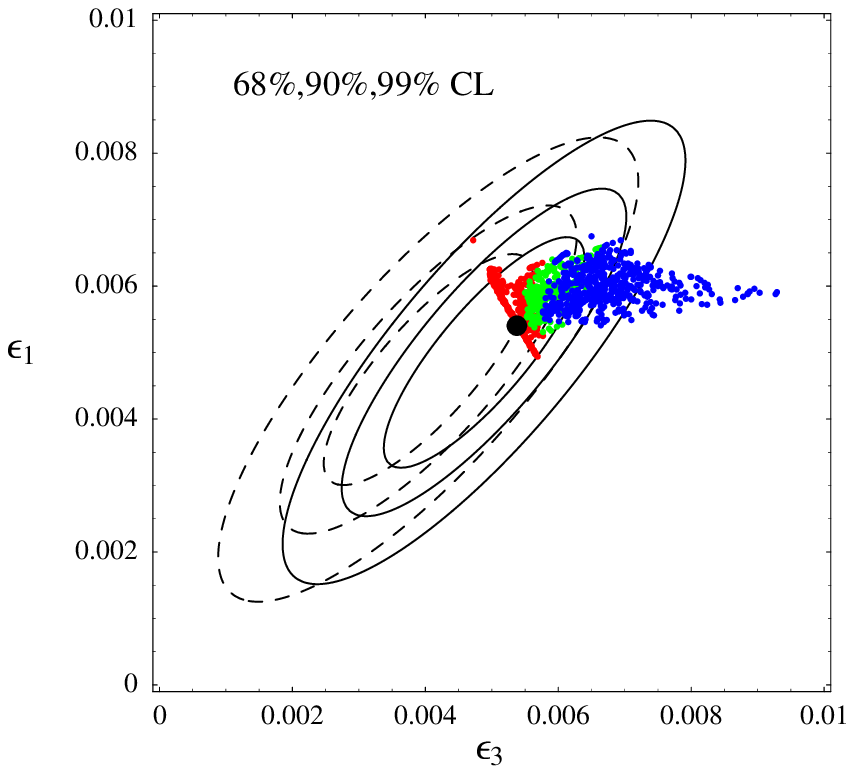,width=0.498\linewidth} \qquad
\epsfig{file=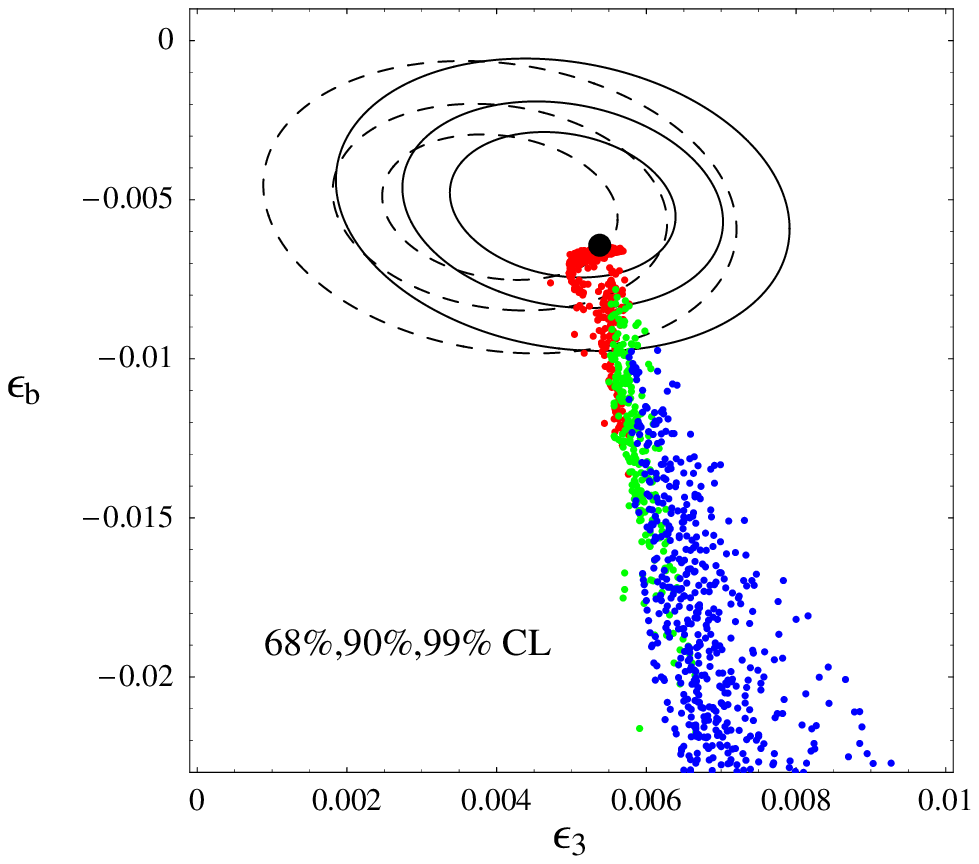,width=0.498\linewidth}
\end{minipage}
\caption{\it Contour plots in the planes $(\e{3},\e{1})$ and $(\e{3},\e{b})$
as obtained from eq.(\ref{eq:fit}). The dashed contours are obtained by omitting
$A_{FB}^b$ from the fit. Superimposed are the points in the MCHM 
with $N\geq 3$ and $|\delta g_{Lb}/g_{Lb}^{SM}| \leq 2\%$.
Blue (dark) dots correspond to $\eps > 0.3$, green (light) dots to $0.2 < \eps < 0.3$, 
red (medium dark) dots to $\eps < 0.2$. The fat black dot represents the 
SM prediction for $m_\text{Higgs}=115\,\text{GeV}$.}
\label{fig:contours}
\end{figure}

In order to compare with the experimental results (\ref{eq:fit}), we use eqs.(\ref{eq:pred}) 
to convert our prediction for $S$, $\Delta\rho$, $\delta g_{Lb}$ into one for
the $\e{i}$, and then perform a $\chi^2$ test.
We keep only points that have $N\geq 3$ and  satisfy $\Delta\chi^2 \equiv \chi^2-\chi^2_{min} < 13.28$, 
the latter condition corresponding to a $99\%$ CL for a $\chi^2$ with 4 degrees of 
freedom.~\footnote{The $\chi^2$ function is defined as:
\begin{equation}
\chi^2 = \sum_{ij}\, (\e{i}-\mu_i) \left(\sigma^2\right)^{-1}_{ij} (\e{j}-\mu_j)\, ,
 \qquad \left(\sigma^2\right)_{ij} = \sigma_i \rho_{ij} \sigma_j\, .
\end{equation}
where $\mu_i$, $\sigma_i$, $\rho_{ij}$ are respectively the mean values, the standard deviations
and the correlation matrix of eq.(\ref{eq:fit}).}
The results are summarized in Fig.~\ref{fig:full}.~\footnote{Our scatter plots 
are generated using a Mathematica code and have a 
limited number of points due to the limited available CPU time.}
Only a few points with $\eps \gtrsim 0.25$ survive the $\chi^2$ test
(in the plot on the left the points which pass the $\chi^2$ test are the black dots), 
and in any case $\eps$ is never 
larger than~0.3. The amount of fine tuning implied for the minimal model
by the EWPT seems thus to be slightly worse than what was hoped for in Ref.~\cite{Agashe:2004rs}.
This is mainly due to the constraint coming from $Z\to b\bar b$, which
turns out to be actually more stringent than usually assumed in the literature,
Ref.~\cite{Agashe:2004rs} included.
To demonstrate this point, we have repeated the test, now setting
$\e{b}$ to its SM value in the $\chi^2$ function, but at the same time requiring 
$|\delta g_{Lb}/g_{Lb}^{SM}| \leq 1\%$, with $g_{Lb}^{SM}= -0.421$.
The points satisfying these requirements are shown in grey in the left plot of Fig.~\ref{fig:full}.
One can see that by relaxing the constraint from $Z\to b\bar b$, i.e., allowing 
the shift in $g_{Lb}$ to be $1\%$ of the SM value (this is the constraint adopted in 
the NDA estimate of Ref.~\cite{Agashe:2004rs}),~\footnote{As discussed later on in the text, 
the actual constraint is roughly $\delta g_{Lb}/|g_{Lb}^{SM}|\lesssim 0.25\%$, see Fig.~\ref{fig:contours}.}
points with $\eps$ as large
as $\sim 0.5$ are allowed. Notice that these points do pass the test on $S$ and $\Delta\rho$. 
In other words, the strongest constraint on the minimal composite Higgs model 
seems to come from $Z\to b\bar b$, rather than from universal effects.

To better visualize the relative importance of the various $\e{i}$
in constraining the model, we have shown in Fig.~\ref{fig:contours}
the confidence level contours in the planes $(\e{3},\e{1})$, $(\e{3},\e{b})$.
Each plot is obtained by minimizing the $\chi^2$ with respect to the remaining two epsilon 
parameters and computing the $68\%, 90\%, 99\%$ CL contours with 2 degrees of 
freedom.~\footnote{We thank Alessandro Strumia for a clarifying discussion about this point.}
Superimposed on these are the predictions of our model, where this time we have 
shown the set of points which satisfy a preliminary cut of $|\delta g_{Lb}/g_{Lb}^{SM}| \leq 2\%$.
From Fig.~\ref{fig:contours} it is evident that, while $\e{1}$ and $\e{3}$
do not imply a strong tuning in the minimal model, the constraint on
$\e{b}$ requires $\delta g_{Lb}/|g_{Lb}^{SM}| \lesssim 0.25\%$, and
selects points with~$\eps \lesssim 0.25$.

Since the constraint from $\delta g_{Lb}$ is so important, one might ask
how crucial is the inclusion, in the set of experimental observables,
of a measurement like the LEP/SLD Forward-Backward 
asymmetry $A_{FB}^{b}$, that appears 
to deviate by almost 3 sigmas from its SM prediction (see~\cite{LEPEWWG}). 
For example, if this anomaly is a statistical 
fluctuation, and if it were $A_{FB}^{b}$ that mainly 
determines the constraint on $\e{b}$, 
then  this constraint from $\delta g_{Lb}$ could be artificially too restrictive.
To prove that the anomaly does not actually play an important role in the analysis,
we have shown in Fig.~\ref{fig:contours} (dashed lines) the contours one obtains by 
omitting $A_{FB}^{b}$ 
from the fit: while the $\chi^2$ global minimum is considerably lower, the constraint
on $\e{b}$ does not essentially change.~\footnote{Notice also that the best fit
prefers smaller values of $\e{3}$, and this in turn implies a smaller $m_H$. 
The fact that the SM fit prefers (much) smaller Higgs masses if one omits the hadronic 
asymmetries is well known.}
The point is that $R_b$, ratio of the
$b$-quark partial width of the $Z$ to its total hadronic partial width, 
is more sensitive to $g_{Lb}$ than $A^b_{ FB }$, since $g_{Lb}\gg g_{Rb}$:
\begin{align} \label{Rb}
A_{FB}^{b} &\propto \frac{g_{Lb}^2 - g_{Rb}^2}{g_{Lb}^2 + g_{Rb}^2}
\simeq 1 - 2\, \frac{g_{Rb}^2}{g_{Lb}^2}\, , \\[0.2cm]
R_b &\propto g_{Lb}^2 + g_{Rb}^2 \, .
\end{align}
The experimental error in both observables is of the same order (roughly few per mil). 
It is then clear that the most important constraint on $g_{Lb}$ comes from $R_b$,
and hence omitting $A^b_{ FB }$ does not significantly change the constraint on $\epsilon_b$.
On the other hand, $R_b$ is not anomalous and there is no sound reason to omit it from the fit.
Moreover, even inclusive hadronic observables like $\Gamma_Z$, $\sigma_h$, $R_h$,
already pose a significant constraint on $g_{Lb}$: 
they are measured with slightly more precision (per mil) than $R_b$, 
but are clearly less sensitive to $g_{Lb}$.
The resulting constraint on $g_{Lb}$ is roughly similar from both
inclusive observables and $R_b$.
Thus, even omitting all $b$-quark observables from the fit, we do not expect that corrections
to $g_{Lb}$ as large as $1\%$ of its SM value would be allowed.
This proves that our conclusions are robust in this respect.

\section{Discussion: a closer look at the results}
\label{sec:closerlook}

\begin{figure}
\centering
\epsfig{file=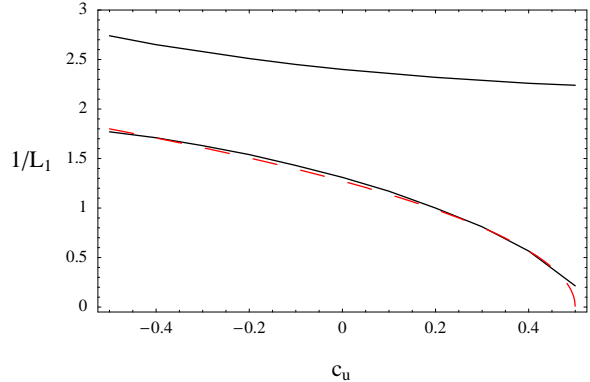,width=0.5\linewidth} 
\caption{\it Mass of the first (lower solid curve) and second (upper solid curve)
$b^\prime$ KK states as functions of $c_u$, in units of $1/L_1 = (v/\eps)\, 2\pi/\sqrt{N}$. 
The red dashed line corresponds to $y(c_u)=1.8\,\sqrt{1/2-c_u}$.
The curves are obtained for a particular choice of the 5D input parameters. Scanning over
the 5D parameters as described in Section~\ref{sec:analysis} leads to a theoretical
uncertainty of order~$\sim~20\%$.}
\label{fig:mbprime}
\end{figure}

The importance of the constraint from $Z\to b\bar b$ could have been anticipated 
from the NDA estimate of Ref.~\cite{Agashe:2004rs},  requiring 
$\delta g_{Lb}/|g_{Lb}^{SM}| \lesssim 0.25\%$.
However, the situation is slightly worse than one might expect from a naive estimate,
in that $\delta g_{Lb}$ is enhanced by the exchange of a state $b^\prime$ 
which becomes light in the limit $c_u \to 1/2$~\cite{Agashe:2004rs}.
The presence of such state is quite a general feature of models
where the strong sector has a custodial symmetry, $b_R^\prime$ being the partner 
under SU(2)$_R$ of the composite state which mixes with the elementary $t_R$.
What is not general, but peculiar to models with a 5D description,
is that the mass of $b^\prime$
is related to the strength of the coupling
of the elementary $t_R$ to the CFT: $m_{b'}\sim m_{\rho}\sqrt{1/2-c_u}$, 
see Fig.~\ref{fig:mbprime}.
This is a consequence of the dual holographic descriptions of 5D theories~\cite{Contino:2004vy}:
in the left-handed description of the bulk field $\xi_u$, $m_{b'}$
comes from the marriage of a composite $b^\prime_R$ with an elementary
$b^\prime_L$ whose coupling to the CFT goes like $\sim\sqrt{1/2-c_u}$~\cite{Agashe:2004rs}.

A more refined NDA estimate for $\delta g_{Lb}$ thus reads~\footnote{In the
limit $c_q\to 1/2$ the factor $(1/2-c_q)$ in eq.(\ref{NDAdg}) should be replaced by 
$1/(2\log(L_1/L_0))$.
For the values of $c_q$ considered in our analysis, $0.36 \leq c_q \leq 0.45$,
the NDA estimate~(\ref{NDAdg}) accurately reproduces the $c_q$ dependence of $\delta g_{Lb}$.
The same consideration also applies to subsequent NDA estimates.}:
\begin{equation} \label{NDAdg}
\frac{\delta g_{Lb}}{g_{Lb}^{SM}} = \eps^2 \left(\frac{1}{2}-c_q \right) F\, ,
\end{equation}
where $F=F(\m,\M,c_q,c_u)$ is expected to reproduce the enhancement due to the light $b'$ state,
$F\sim -1/(1/2-c_u)$ for $c_u\to 1/2$, and have only a mild dependence upon $\m$, $\M$, $c_q$.
As explained in Ref.~\cite{Agashe:2004rs}, it is crucial to consider the correlation of 
$\delta g_{Lb}$ with $m_t$, whose NDA estimates is:
\begin{equation} \label{NDAmt}
m_t = \frac{4\pi}{\sqrt{N}}\, v 
 \sqrt{\left(\frac{1}{2}-c_q\right) \left(\frac{1}{2}+c_u\right)}\, \eta \, .
\end{equation}
Here $\eta$ is an $O(1)$ function of $\m$, $\M$, $c_q$ and $c_u$ that parametrizes a possible deviation
from strict NDA. 
Eqs.(\ref{NDAdg}) and (\ref{NDAmt}) show the tension between $\delta g_{Lb}$ and $m_t$:
it is not possible to suppress the correction to $Z\to b\bar b$ (unless tuning $\eps$ to 
small values), without making at the same time
the top Yukawa coupling unacceptably small.
The tension arising from the dependence of $m_t$ and $\delta g_{Lb}$ on $c_q$
is well-known (it was first discussed in~\cite{Agashe:2003zs}), but that due to 
$c_u$ ($m_t$ and $\delta g_{Lb}$ prefer $c_u \rightarrow \pm 1/2$, respectively),
is a consequence of the light $b^\prime$, and it has not been discussed in detail before.
Were it not for the $c_u$ dependence $F\sim -1/(1/2-c_u)$ in eq.(\ref{NDAdg}), 
one could reproduce the experimental value of $m_t$ for smaller values of $(1/2-c_q)$ 
by letting $c_u\to +1/2$, thus reducing the size of $\delta g_{Lb}$.
\begin{figure}[t]
\centering
\epsfig{file=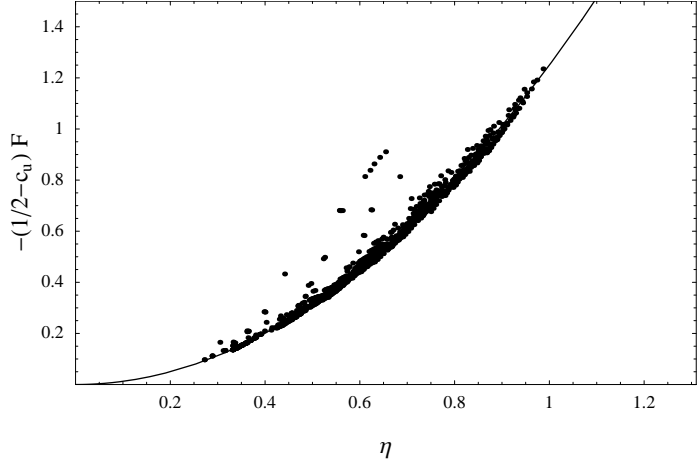,width=0.6\linewidth} 
\caption{\it Correlation between $\delta g_{Lb}$ and $m_t$:
scatter plot in the plane $(\eta,-(1/2-c_u) F)$ 
for the set of points 
with $N\geq 3$ and $|\delta g_{Lb}/g_{Lb}^{SM}| \leq 2\%$.
Here $\eta$ is defined by eq.(\ref{NDAmt}) and $F$ by eq.(\ref{NDAdg}).
The solid line corresponds to the curve $y(\eta)= 1.25\, \eta^2$.}
\label{fig:dgLbNDA}
\end{figure}

Analyzing the set of points obtained by scanning over the parameter space of the MCHM,
we found that eq.(\ref{NDAdg}) is quite well reproduced for 
\begin{equation} \label{mtdgcorr}
F \simeq \eta^2 f(c_u)\,
\end{equation}
with $f(c_u) = -1.25/(1/2-c_u)$. This means that, 
in addition to the tension due to $c_u$, $c_q$ discussed above,
$m_t$ and $\delta g_{Lb}$ are 
strongly correlated 
through the function $\eta$. 
This correlation is evident from the scatter plot in Fig.~\ref{fig:dgLbNDA}, 
where we plotted $-(1/2-c_u)F$ as a function of $\eta$ for the sets of points
with $N\geq 3$ and $|\delta g_{Lb}/g_{Lb}^{SM}| \leq 2\%$.
We find $0.3 \lesssim \eta\lesssim 1$.
The fact that the points do not exactly lie on a curve, but there is some spread,
indicates that eq.(\ref{mtdgcorr}) is only an approximation, though quite accurate. 
This in turn implies that it is not possible, and not meaningful indeed, to determine
the functional form of $f(c_u)$ with a very high degree of precision.
In fact, although there is a clear indication of the behavior $f(c_u)\sim -1/(1/2-c_u)$ 
for $c_u\to 1/2$, more general forms, like for example $f(c_u)= A + (B+C\, c_u)/(1/2-c_u)$
with $A$, $B$, $C$ constants,  
also give a good fit, and are actually more theoretically motivated since the 
function $F$ is expected to encode also contributions to $\delta g_{Lb}$ that are not
enhanced by the light $b'$ state, such as that from the Higgs coupled to the vector
resonances. 
The uncertainty on the form of $F$ thus implies that the different contributions to $\delta g_{Lb}$
cannot be easily disentangled in our calculation.

\begin{figure}
\centering
\begin{minipage}[t]{0.95\linewidth}
\centering
\epsfig{file=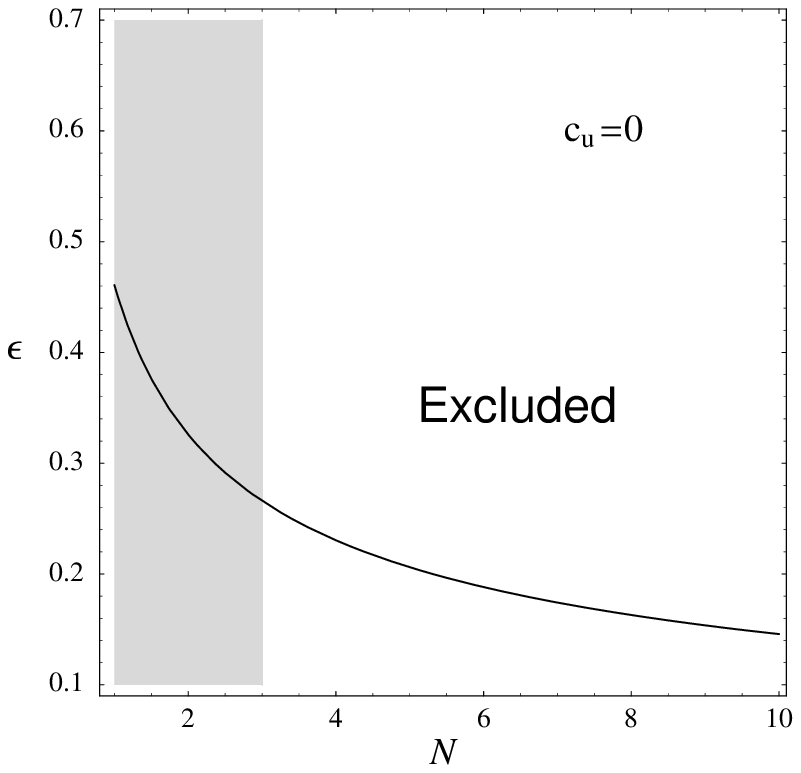,width=0.45\linewidth} \qquad\quad
\epsfig{file=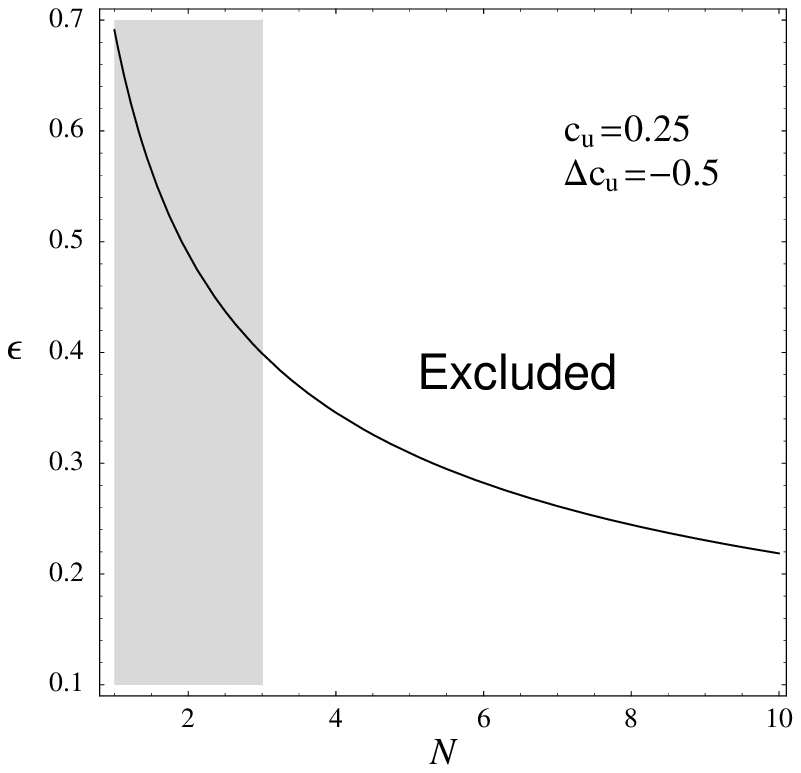,width=0.45\linewidth}
\end{minipage}
\caption{\it Left plot: Constraint $\delta g_{Lb}/|g^{SM}_{Lb}|<0.25\%$ in the plane $(N,\eps)$ 
from eq.(\ref{dgcontour}), after optimizing $c_u=0$. 
Right plot: same as before but
allowing a splitting $\Delta c_u=-0.5$ in $\xi_u$ (see text), and optimizing $c_u=0.25$.
In the grey region $N<3$ and the $1/N$ perturbative expansion cannot be trusted.}
\label{fig:epsN}
\end{figure}

The strong correlation between $m_t$ and $\delta g_{Lb}$
through the function $\eta$
can be explained by assuming that the main dependence of $\eta$ upon 
$\m$, $\M$, $c_q$ and $c_u$ comes from 
the mixing between elementary and composite fields.
In particular, it implies that
the value of $\delta g_{Lb}$ in our model is
completely determined, as a function of $N$, $\eps$ and $c_u$,
once we fix $m_t$ to its experimental value. Indeed, by extracting $\eta$ from
eq.(\ref{NDAmt}) and setting 
$F = -1.25\, \eta^2/(1/2-c_u)$
in eq.(\ref{NDAdg}), one obtains
\begin{equation} \label{dgcontour}
\frac{\delta g_{Lb}}{g_{Lb}^{SM}} \simeq 
- 0.46\, \eps^2\,
 \frac{N}{16 \pi^2} \left(\frac{1}{4}-c_u^2 \right)^{-1}
 \left(\frac{m_t^{\overline{\text{MS}}}(2\,\text{TeV})}{150\, \text{GeV}}\right)^2\, .
\end{equation}
Thus, $\delta g_{Lb}$ is minimized for $c_u=0$. Fig.~\ref{fig:epsN} (left plot) shows the constraint
$\delta g_{Lb}/|g_{Lb}^{SM}| <0.25\%$ in the plane $(N,\eps)$, coming from eq.(\ref{dgcontour}), 
after optimizing $c_u$, i.e., setting it to $0$. 
For $N\geq 3$, $\eps$ is bounded to be roughly smaller than $0.25-0.3$, consistently
with what found in the previous section.
This constraint can be relaxed by considering values $N<3$, a limit, however, 
in which the $1/N$ perturbative expansion becomes questionable.

Another way to allow larger values of $\eps$ would be by turning on a small
breaking of the custodial symmetry in the CFT sector, in order to 
break the correlation between the mass of the $b^\prime$ and $m_t$.
For example, a splitting $\Delta c_u$ between the bulk mass of the upper and lower
SU(2)$_R$ components of $Q^u$ inside $\xi_u$ has the effect of
shifting $c_u\to c_u+\Delta c_u$ in eq.(\ref{NDAdg}), without altering eq.(\ref{NDAmt}).
As a consequence, the factor $(1/4-c_u^2)^{-1}$ in eq.(\ref{dgcontour}) is replaced 
by $[(1/2-c_u-\Delta c_u)(1/2+c_u)]^{-1}$.
For example, if $\Delta c_u =-0.5$, then $\delta g_{Lb}$ is minimized for $c_u=0.25$ and the constraint
in the plane $(N,\eps)$ is that shown in the right plot of Fig.~\ref{fig:epsN}.
Although the bound is still quite restrictive, values of $\eps$ as large as $0.4$ are now
allowed for $N\geq 3$.
A small breaking of custodial symmetry is therefore an interesting possibility
to reduce the fine tuning of the minimal model.~\footnote{For example, it could be shined 
from the UV brane into the bulk by the profile of some 5D scalar field.}
The additional contribution to the $\rho$ parameter, 
$\Delta\rho\sim \epsilon^2 \, ( \Delta c )^2 \, x$,
can be under control as long as $\Delta c_u \lesssim 1/2$ and $x$ is somewhat smaller than 1,
where $x$ parametrizes the ratio of the strength of custodial isospin breaking
in the gauge sector relative to the fermion sector 
(see Ref.~\cite{Agashe:2003zs} for a specific example where  $x \sim 1/4$).
At the same time, one should check that the specific realization of the breaking
does not lead to a dangerous extra contribution to the Higgs potential.

At this point, we would like to comment on the sign of $\delta g_{Lb}$.
It turns out that a negative value of $\delta g_{ L b }$ (which
corresponds to $\delta\epsilon_b > 0$) is less
constrained by the data, as one can see from Fig.~\ref{fig:contours}. 
However, {\em both} the contribution to $\delta g_{ L b }$
coming from diagrams with the Higgs coupled to vector resonances
(``gauge'' contribution), and that enhanced by the light $b'$ state
are positive as follows.
The gauge contribution was computed in Ref.~\cite{Agashe:2003zs} for the case
of the Higgs localized on the IR brane, and shown to be positive.
We do not expect that the profile of $A_5$ or the different realization of
fermions in the bulk (as long as $-1/2 < c_q, c_u< +1/2$) can change the sign.
The second contribution
comes from the mixing of $b_L$ with an $SU(2)_L$ singlet $b^{ \prime }$.
Since the sign of the couplings of $b_L$ and $b^{ \prime }$ to $Z$
are opposite, it is easy to check that the sign of the shift in 
the coupling of the physical left-handed bottom is again positive.
Of course, if $b_L$ were to mix with a massive $b^{ \prime }$ state 
with different electroweak quantum numbers, such as an SU(2)$_L$ triplet,
then a negative shift would be
possible.~\footnote{We thank John March-Russell and Riccardo Rattazzi for
having pointed this to us.}

Another interesting possibility is that of a shift in the coupling $g_{Rb}$ of $b_R$ to $Z$, 
since a $\sim 10 \%$ {\em increase} in the magnitude of $g_{Rb}$ can explain
the $A^b_{ FB }$ anomaly.
In turn, such an effect would allow a positive $\delta g_{ Lb }$ at the $1 \%$ level,
while keeping $R_b$ and the inclusive hadronic observables fixed.
In this way the constraint on our model would be relaxed.
Such a large shift in $g_{Rb}$ could only be obtained if $b_R$ has a 
large coupling to the CFT, which in the 5D picture corresponds to
having the $b_R$ wave function localized near the IR brane~\footnote{Of course, 
one would then lose the elegant explanation
of the $m_t / m_b$ hierarchy in terms of 5D wave-functions.}.
However, just like for the case of $b_L$ above, it is easy to see that
this results instead in a reduction in the magnitude of $g_{Rb}$, 
unless $b_R$ mixes with an exotic 
$b^{ \prime }$ state, such as an $SU(2)_L$ doublet with hypercharge $-5/6$ 
(see for example~\cite{Choudhury:2001hs}). 
Hence, it is not possible to relax the constraint
on $\delta g_{ Lb }$ in this way, at least in the context of the minimal model.

\begin{figure}
\centering
\epsfig{file=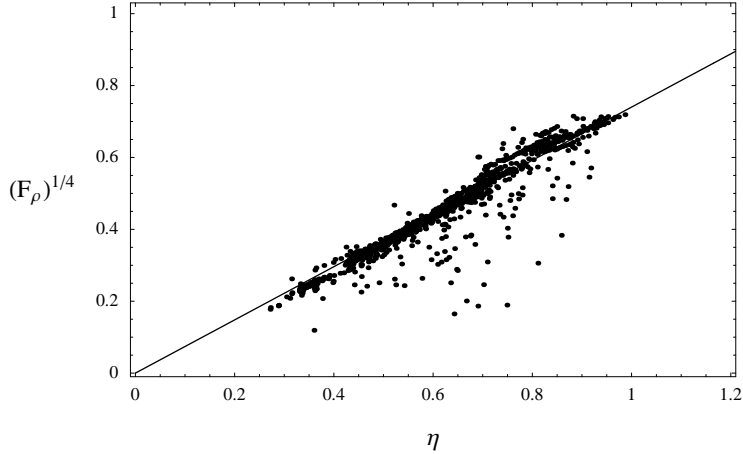,width=0.6\linewidth}
\caption{\it Correlation between $\Delta\rho$ and $m_t$: 
scatter plot in the plane $(\eta,F_\rho^{1/4})$ 
for the set of points 
with $N\geq 3$ and $|\delta g_{Lb}/g_{Lb}^{SM}| \leq 2\%$.
Here $F_\rho$ is defined by eq.(\ref{deltarhoNDA}). 
The solid line corresponds to the curve $y(\eta)= (0.3)^{1/4} \eta$.}
\label{fig:corrdeltarho}
\end{figure}

\begin{figure}
\centering
\epsfig{file=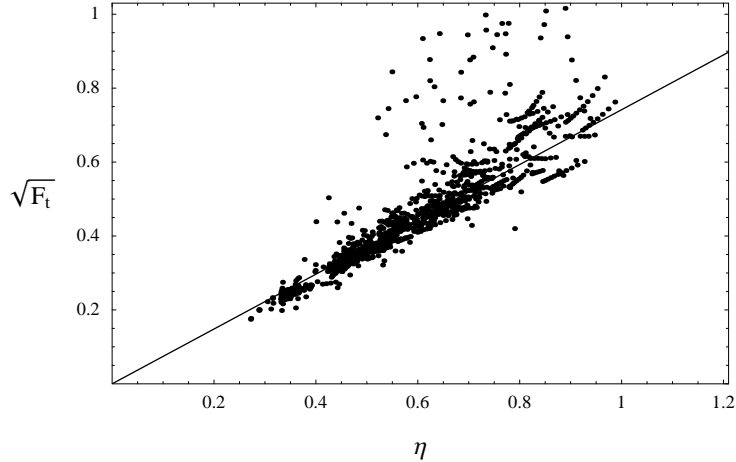,width=0.6\linewidth}
\caption{\it Correlation between $m_\text{Higgs}$ and $m_t$: 
scatter plot in the plane $(\eta,\sqrt{F_t})$, for the set of points 
with $N\geq 3$ and $|\delta g_{Lb}/g_{Lb}^{SM}| \leq 2\%$.
Here $F_t$ is defined by eq.(\ref{mHNDA}) with $F_g=0.20$.
The solid line corresponds to the curve $y(\eta)= \sqrt{0.55}\, \eta$.
}
\label{fig:mHNDA}
\end{figure}

Studying the correlation with $m_t$ is also a useful technique to better understand 
the prediction of the MCHM for $\Delta\rho$ and the Higgs mass.
The NDA estimate for $\Delta\rho$ reads~\cite{Agashe:2004rs}:
\begin{equation} \label{deltarhoNDA}
\Delta\rho = \eps^2\, \frac{N_c}{N} \left(\frac{1}{2}+c_u\right)^2 F_\rho\, ,
\end{equation}
where $F_\rho$ parametrizes the deviation from strict NDA. 
We find that eq.(\ref{deltarhoNDA}) well reproduces 
the set of points for $F_\rho \simeq 0.3\, \eta^4$,
see Fig.~\ref{fig:corrdeltarho}.
Thus, although not as strong as in the case of $\delta g_{Lb}$, 
there is evidence for a correlation also between $\Delta\rho$ and $m_t$.
Moreover, the factor $F_\rho\sim \eta^4$ implies a considerable suppression in 
$\Delta\rho$ compared to the naive estimate, since $0.3\lesssim \eta\lesssim 1$.
The NDA estimate for the physical Higgs mass consists of a contribution 
from the top quark and one from the gauge fields~\cite{Agashe:2004rs}:
\begin{equation} \label{mHNDA}
\begin{split}
m_\text{Higgs}^2 
 &= 2 v^2 \left[ F_t\, \frac{N_c}{N^2}\, 16\pi^2 \left(\frac{1}{2}-c_q\right)
    \left(\frac{1}{2}+c_u\right) - F_g\, \frac{3 g^2}{N}\right]\, , \\
 &= 2 \, \frac{N_c}{N}\, \frac{F_t}{\eta^2}\, m_t^2- 2 v^2\, F_g\, \frac{3 g^2}{N}\, .
\end{split}
\end{equation}
As before, $F_t$, $F_g$ parametrize the deviation from strict NDA
of the top and gauge contributions respectively. While $F_g$ is just an $O(1)$ constant,
$F_t$ is expected to have a (mild) dependence upon $\m$, $\M$, $c_q$, $c_u$.
We find that eq.(\ref{mHNDA}) gives a good description of 
our set of points for $F_t\simeq 0.55\, \eta^2$, $F_g\simeq 0.20$,
see Fig.~\ref{fig:mHNDA}.

\section{Spectrum of new particles}
\label{sec:spectrum}

Determining the spectrum of new particles is of extreme importance 
for studying the phenomenology of our model at future colliders.
For this reason we present here a detailed analysis of the spectrum
of vectors and fermions, including the effects of electroweak symmetry breaking.

According to the holographic description,
the masses of the new states can be extracted from the
poles or the zeros of the two-point form factors.
As discussed in Ref.~\cite{Contino:2004vy}, the case of 
one elementary source $\Psi$ coupled to one tower of CFT resonances is quite
straightforward. By integrating out the CFT states, one can derive the effective action
for the source at the quadratic level:
\begin{equation} \label{eq:Leff}
{\cal L}_{eff} = \bar\Psi \Sigma(p) \Psi \, .
\end{equation}
The two-point function $\Sigma$ encodes all the information about the spectrum.
If the source is non-dynamical, i.e. it is just a probe to excite 
the CFT mesons out of the vacuum, the spectrum of composite states is given by the 
\textit{poles} of~$\Sigma$.
If instead the source is dynamical, it mixes with the tower of composite states 
and distorts their spectrum.
The resulting eigenstates are partially composite modes, and their masses
are given by the \textit{zeros} of~$\Sigma$.

Extracting the spectrum when one or more sources couple to several towers mixed with 
each other
(as it happens, for example, due to the IR-brane mass mixing terms or to the Higgs vev),
is only slightly more complicated. Consider, for example, the case of two
sources $\Psi_{1,2}$ coupled to two mixed towers of CFT resonances.
By integrating out the CFT states, the effective action  will have the
form (\ref{eq:Leff}) with $\Psi = (\Psi_1 ,\Psi_2)$ and
\begin{equation}
\Sigma(p) = \begin{pmatrix} M_{11}(p) & M_{12}(p) \\ M_{12}(p) & M_{22}(p) \end{pmatrix} \, .
\end{equation}
If both $\Psi_{1}$ and $\Psi_{2}$ are dynamical, the spectrum is clearly given by the
zeros of the determinant of $\Sigma$, as one can simply 
determine by rotating to the
basis in which there are two orthogonal CFT towers, each coupled to one
elementary source. If instead both sources are non-dynamical, then the full spectrum
of composite states is given by the poles of \textit{any} of the entries of $\Sigma$.
This is because it does not matter which source excites the mesons out of the
vacuum, as long as the latter are mixed.
Finally, when only one source is dynamical, say $\Psi_1$, 
the physical spectrum of KK states is given by
the zeros of $M_{11}$. Indeed, although $\Psi_1$ is directly coupled only
to the first tower, it can probe the full spectrum, since all CFT states are mixed
with each other.

By applying these rules, one can easily extract the KK spectrum of the MCHM.
Before EWSB, the resonances of the strong sector come in four vectorial and 
three fermionic KK towers. The lightest mass of each tower can be conveniently 
expressed in terms of the typical mass $m_\rho \equiv 4\pi/\sqrt{N}\, f_\pi$ expected from NDA:
\begin{equation} \label{mKK}
M_i = c_i \, m_\rho \, , \qquad\quad
 m_\rho = \frac{4\pi}{\sqrt{N}}\, \frac{v}{\eps}\, .
\end{equation}
In the vector case, the spectrum is given by:
\begin{itemize}
\item[--] a tower of $W$'s (\textbf{3}$_{0}$ of SU(2)$_L\times$U(1)$_Y$) 
with masses given by: 
 zeros$\displaystyle{\left\{ \Pi_0 -\frac{p^2}{g^{2}_{UV}} \right\}}$. \\
The lightest eigenvalue is of the form (\ref{mKK}) with
\begin{equation} \label{cW}
c_W \simeq \frac{x_0}{2} \left(1+\frac{1}{2}\, \frac{g^2}{g_5^2 k}\, r(x_0) \right)\, ,
\end{equation}
where $r(x)=\pi/x\, (Y_0(x)- z_{IR}\, x\, Y_1(x))/(J_1(x)+ z_{IR}\, x\, J_0(x))$ and 
\begin{equation}
x_0(z_{IR}) \simeq \frac{2.4}{\sqrt{1+9\pi^2/32\, z_{IR}^2}}
\end{equation}
corresponds to the first zero of $\tilde J_0(x) = J_0(x) - z_{IR}\, x\, J_1(x)$,
$J_{0,1}$ and $Y_{0,1}$ being Bessel functions. Here $g$ is the SU(2)$_L$ low-energy gauge coupling,
$z_{IR}=g_5^2 k/g_{IR}^2$, and
$1/g_5^2$, $1/g_{UV}^2$, $1/g_{IR}^2$ denote,
respectively, the coefficients of the gauge kinetic term for SO(5) in the bulk, 
SU(2)$_L$ on the UV brane, SO(4) on the IR brane.  
\item[--] a tower of $Y$'s (\textbf{1}$_{0}$ of SU(2)$_L\times$U(1)$_Y$) 
with masses given by: 
zeros$\displaystyle{\left\{ \Pi_0+\Pi_0^B -\frac{p^2}{g^{\prime\, 2}_{UV}} \right\}}$. \\[0.1cm]
Here $1/g^{\prime\, 2}_{UV}$ stands for the coefficient of the U(1)$_Y$ kinetic term on the UV brane.
If $z_{IR}^B \lesssim z_{IR}$, where $z_{IR}^B=g_{5 B}^{2} k/(g_{IR}^{B})^2$ denotes
the ratio between the IR-brane and bulk kinetic term of U(1)$_{B-L}$, then the
lightest state has $c_Y$ as given by eq.(\ref{cW}) with $g$ replaced by the low-energy 
hypercharge~$g'$.
In the opposite limit $z_{IR}^B \gtrsim z_{IR}$, the first KK mode of the $Y$ tower 
is lighter than that of the $W$ tower: $c_Y$ is still approximately given by eq.(\ref{cW})
with $g\to g'$, provided we also replace $g_5$ ($g_{IR}$) with $g_{5 B}$ ($g_{IR}^B)$.
\item[--] a tower of $\hat W$'s (\textbf{2}$_{1/2}$ of SU(2)$_L\times$U(1)$_Y$) 
with masses given by: poles$\displaystyle{\left\{\Pi_0+ \frac{1}{2}\Pi_1 \right\}}$.\\
The lightest eigenvalue is of the form (\ref{mKK}) 
with $c_{\hat W} = 1.96 \simeq 5\pi/8$.
\item[--] a tower of $X$'s (\textbf{1}$_{\pm 1}$ of SU(2)$_L\times$U(1)$_Y$) 
with masses given by: poles$\displaystyle{\left\{\Pi_0 \right\}}$. \\
The lightest eigenvalue has $c_{X} = x_0/2$.
\end{itemize}
From the holographic viewpoint, the $W$ and $Y$ states are partially composites: 
their masses are given by the zeros of two-point functions, and the lightest mass (\ref{cW})
has the form one would expect for a pure CFT eigenvalue distorted by a small 
perturbation $\propto g^2$. The states $\hat W$ and $X$ are, on the contrary, pure composites.

The fermionic spectrum consists, before EWSB, of three towers:
\begin{itemize}
\item[--] a tower of $q_L$'s (\textbf{2}$_{1/6}$ of SU(2)$_L\times$U(1)$_Y$) 
with masses given by: zeros$\displaystyle{\left\{\pslash \left(F_0^q+F_1^q \right)\right\}}$. \\
The lightest mass, as obtained by scanning over the parameter space in the way 
described in Section~\ref{sec:analysis}, has the form (\ref{cW}) with $c_{q_L} \simeq 0.49 - 0.85$.
\item[--] a tower of $t_R$'s (\textbf{1}$_{2/3}$ of SU(2)$_L\times$U(1)$_Y$) 
with masses given by: zeros$\displaystyle{\left\{\pslash \left(F_0^u - F_1^u \right)\right\}}$. \\
The lightest mass has $c_{t_R}\simeq 0.95 - 1.1$.
\item[--] a tower of $b_R$'s (\textbf{1}$_{-1/3}$ of SU(2)$_L\times$U(1)$_Y$) 
with masses given by: poles$\displaystyle{\left\{\pslash \left(F_0^u - F_1^u \right)\right\}}$. \\
The lightest mass has $c_{b_R}\simeq (0.9 - 1.1) \sqrt{1/2-c_u}$, where the presence
of the extra factor $\sqrt{1/2-c_u}$ has been discussed in the previous section.
\end{itemize}

\begin{figure}[t]
\begin{minipage}[t]{0.95\linewidth}
 \epsfig{file=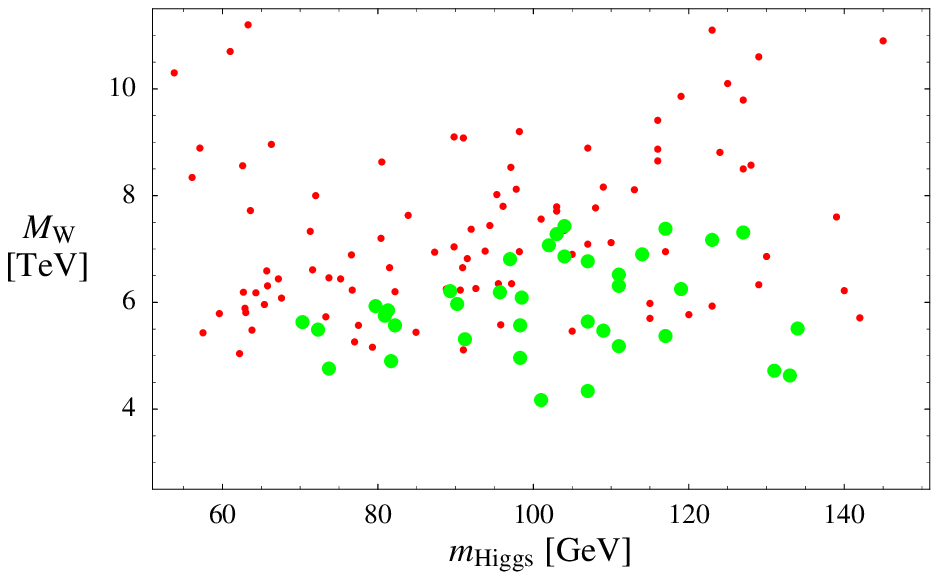,width=0.49\linewidth} \qquad
 \epsfig{file=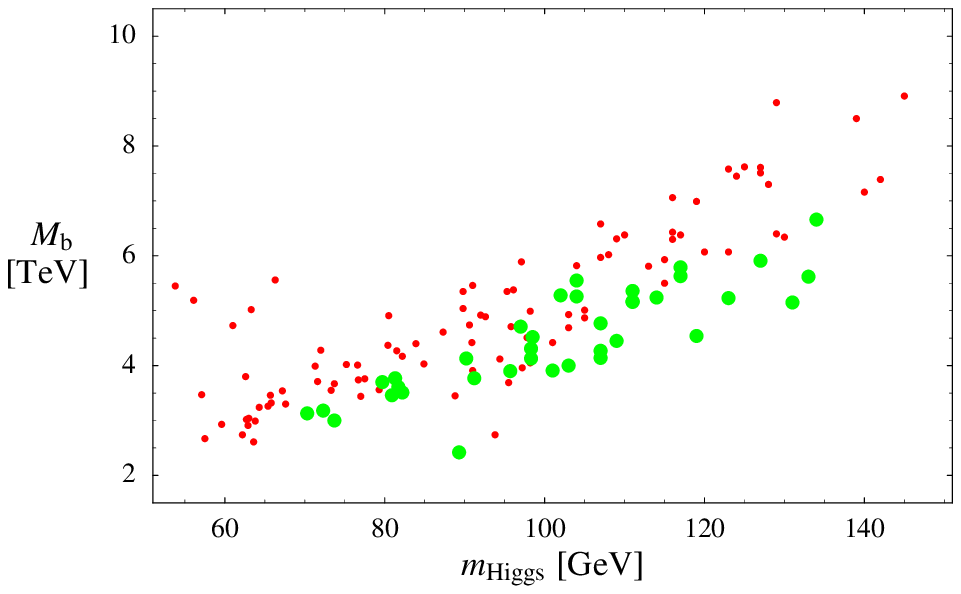,width=0.49\linewidth}
\end{minipage}
\caption{\it Masses of the lightest $W$ and $b$ KK states for the set of points 
that satisfy the $\chi^2$ test ($\Delta\chi^2 < 13.28$, $N\geq 3$).
Green fat dots correspond to $0.2<\eps<0.4$, small red dots to $\eps<0.2$.
}
\label{fig:KKmasses}
\end{figure}

After EWSB, different KK towers are mixed by the Higgs vev.
The final spectrum consists of:
\begin{itemize}
\item[--] a tower of charged vectors ($W$'s) with masses given by: 
 zeros$\displaystyle{\left\{ \Pi_0 + \frac{\eps^2}{4}\Pi_1 - %
       \frac{ p^2}{ g^{2}_{UV}} \right\}}$.
\item[--] two towers of neutral vectors ($Z$'s) with masses given by: \\[0.2cm] 
 \hspace*{0.1cm}
 zeros$\displaystyle{\left\{ \left(\Pi_0 -\frac{p^2}{g^{2}_{UV}}\right) %
       \left(\Pi_0+\Pi_0^B -\frac{p^2}{g^{\prime\, 2}_{UV}}\right) %
       + \frac{\eps^2}{4} \Pi_1 \left[ \Pi_0^B + 2\Pi_0 - p^2 %
         \left( \frac{1}{g^{2}_{UV}} + \frac{1}{g^{\prime\, 2}_{UV}}\right) \right] \right\}}$\\[0.2cm]
 \hspace*{0.1cm}
 poles$\displaystyle{\left\{\Pi_0+ \frac{1}{2}\Pi_1 \right\}}$.
\item[--] a tower of charge $+2/3$ fermions ($t$'s) with masses given by: \\[0.2cm] \hspace*{0.1cm}
 zeros$\displaystyle{\left\{ p^2 \left(F_0^q+F_1^q \sqrt{1-\eps^2}\right)
   \left(F_0^u - F_1^u \sqrt{1-\eps^2}\right) - \eps^2 (M_1^u)^2 \right\}}$.
\item[--] a tower of charge $-1/3$ fermions ($b$'s) with masses given by:
 zeros$\displaystyle{\left\{ \pslash \left(F_0^q+F_1^q \sqrt{1-\eps^2}\right) \right\}}$. 
\end{itemize}
Figure~\ref{fig:KKmasses} shows the spectrum of the lightest $W$ and $b$ KK
states, obtained using the above formulas
for the set of points that satisfy the $\chi^2$ test.
For $m_\text{Higgs}\sim 115\,$GeV, both $W$ and $b$ 
states can be as light as $\sim 4$ TeV,
although this prediction can be slightly modified by varying the size of the IR-brane
kinetic terms.
In the case of the fermions, the IR-brane terms for $\xi_u$ and $\xi_q$
have been set to zero for simplicity in our analysis.
Turning them on will make the spectrum lighter, although we expect $Z\to b\bar b$ 
to impose a strong constraint on their values.
In the case of the spectrum of $W$'s, the left plot of
Fig.~\ref{fig:KKmasses} follows from having chosen the SO(4) kinetic term on the IR 
brane to vary between $1/16\pi^2 \leq 1/g_{IR}^2 \leq 3/16\pi^2$.
Larger values of the SO(4) IR term are not strongly constrained by the $\chi^2$ test
due to the much more restrictive bound from $Z\to b\bar b$, and they would imply lighter $W$'s.
In the same way, the spectrum of $Z$'s depends on the IR-brane kinetic
term for U(1)$_{B-L}$. The latter is even less constrained by
the electroweak fit, and this leaves open the possibility of a lighter spectrum of
$Z$ resonances, though this is not required, i.e. it is not
a prediction of the model.
For example, $z^B_{IR}=2\, z_{IR}^2$ ($z^B_{IR}=4\, z_{IR}^2$) gives a spectrum of $Z$'s
$\sim 15\%$ ($30\%$) lighter than that of the $W$'s, while in the opposite limit 
$z_{IR}^B < z_{IR}$ the $Z$'s are the heaviest vectorial states.
Finally, we found that the lightest $t$ states are almost degenerate with the $b$'s.

\section{Conclusions}

The complete analysis of electroweak precision observables that we have performed
in this paper
for the minimal composite Higgs model of Ref.~\cite{Agashe:2004rs} 
showed that the strongest constraint in the fit is set by $Z\to b\bar b$ data.
These can be reproduced for $\eps = v/f_\pi$ of order 0.25 or smaller,
which requires a tuning in the Higgs potential at the level of a few percent.
Corrections encoded by the $S$ parameter and $\Delta\rho$, on the other hand, 
are compatible with current data for values of $\eps$ as large as 0.5 and
do not imply a significant amount of tuning 
in the parameters of the model.
Effects from four fermions operators 
(which can be encoded in the $W$, $Y$ parameters
of Ref.~\cite{Barbieri:2004qk}) are small 
and do not impose any relevant constraint.

Although the restrictive constraint from $Z\to b\bar b$ depends on the details
of the specific realization of fermions, and as such it is model dependent
to a certain extent, a simple NDA estimate
shows that the bound is quite general once 
one assumes that the elementary fermions couple linearly to the CFT sector.
At the root of the problem lies the tension between  reproducing
a large enough top quark mass while keeping $\delta g_{Lb}$ small.
In the specific case of the MCHM, we found that the leading correction to
$Z\to b\bar b$ comes from the exchange of a light $b'$ state.
The presence of the $b'$ state is a robust consequence of custodial symmetry
and the way it is realized in our model, where the elementary $t_R$ couples
to a CFT operator transforming as a doublet under SU(2)$_R$.
The naive estimate $\delta g_{Lb}/g_{Lb}^{SM} \sim m_t^2/m_{b'}^2$ which follows
is expected to hold also in other composite Higgs models with a similar realization 
of custodial symmetry.
In the MCHM the constraint is made even stronger, 
since the mass of the $b'$ is 
smaller than the other KK masses and is, in fact, 
tied to the strength of the coupling of the elementary
$t_R$ to the CFT. This in turn implies a strict correlation between $m_t$ and
$Z\to b\bar b$, as a result of which we were able to derive a bound on 
$\eps$ in terms of the number $N$ of CFT colors only, with no further free parameter.
The constraint obtained in this way is similar to that derived from the global
electroweak fit.

A possible way to weaken the constraint from $Z\to b\bar b$ and reduce the level of fine
tuning might be a different realization of the custodial symmetry. 
Given that the Higgs transforms as a real
bidoublet under SU(2)$_L\times$SU(2)$_R$, one can demand that the elementary $t_R$ and $q_L$ 
couple to CFT operators that transform respectively as 
a singlet and a bidoublet under 
SU(2)$_L\times$SU(2)$_R$. 
In this way custodial symmetry does not require a $b'$ partner of $t_R$,
and the consequent tree-level contribution to $Z\to b\bar b$ can be avoided.
This in turn breaks the correlation between $\delta g_{Lb}$ and $m_t$ through the
parameter $c_u$, suggesting that it might be possible to suppress the remaining contributions
to $\delta g_{Lb}$ while keeping $m_t$ fixed to its experimental value.
This different implementation of custodial symmetry is realized, for example, 
in the SU(5)/SO(5) intermediate Higgs model of Ref.~\cite{Katz:2005au}, though 
the issue of $Z\to b\bar b$ is not discussed in that paper.
The same construction can be easily implemented in a 5D setup.
Notice that in this different realization $\Delta\rho$ is also suppressed, 
since we need to exchange the elementary $q_L$ to break custodial symmetry.
In the case of the SU(5)/SO(5) model 
(or even for an SU(2)$_L \times $SU(2)$_R$ scenario where the Higgs is composite but not a PGB), 
we estimate $\Delta\rho\sim (1/2-c_q)^2 \eps^2 N_c/N$.

Alternatively, one could try to avoid the correlation between $\delta g_{Lb}$ and $m_t$
by turning on a small breaking of custodial symmetry in the strong sector. 
We estimated that the extra contribution to $\Delta\rho$ can be under control, although
one should check that the specific realization of the breaking does not lead to 
unwanted extra contributions to the Higgs potential. 
Finally, we showed that a larger shift in $g_{Lb}$ can be accommodated
if $b_L$ or $b_R$ mix with exotic $b^{ \prime }$ states
(i.e. states with electroweak quantum numbers different from those
of a SM bottom quark).
If $b_L$ mixes with an exotic $b^\prime$ state, this can lead to $\delta g_{Lb} < 0$,
which is less constrained by data (see Fig.~\ref{fig:contours}).
On the other hand, a $\sim 10\%$ positive shift in $g_{Rb}$, as coming
from the mixing of $b_R$ with an exotic $b^\prime$, would in turn relax the 
constraint on $g_{Lb}$ from $R_b$, allowing for $\delta g_{Lb}/|g_{Lb}^{SM}|\sim 1\%$.

Instead of relaxing the bound on $\eps$ imposed by $Z\to b\bar b$, one could
try to obtain naturally a small $\eps$ by modifying the minimal model.
Ref.~\cite{Agashe:2004rs} showed that a small deformation of the Higgs potential, 
as might come for example from the 1-loop contribution of heavy bulk  fermions, 
can lead to values $\eps\simeq 0.2-0.4$ with virtually no fine tuning. 
It was also shown that this can help increase the physical Higgs mass
while allowing for larger values of $N$,
thus improving the behaviour of the perturbative expansion.
Indeed, to have  $m_\text{Higgs}\gtrsim 115\,$GeV in the minimal model one is
restricted to consider small numbers of CFT colors, $N\lesssim 4-5$ (see Fig.~\ref{fig:full}), 
so that next-to-leading order corrections in the perturbative
expansion are expected to be large. This very observation, however, implies that
values of $m_\text{Higgs}\sim 100\,$GeV in Fig.~\ref{fig:full} might
not actually be ruled 
out by the LEP direct bound, due to the large theoretical uncertainty.
Moreover, the Higgs mass can be slightly increased even in the context of the MCHM 
by allowing for a larger SO(4) IR-brane kinetic term. In this limit the first
gauge KK state becomes lighter, the gauge correction to the Higgs potential
is cut-off at a smaller scale, and this in turn results 
in a heavier physical Higgs. 

Besides naturalness, maybe a more pressing issue to address is the
potential for discovery of the model at present and future colliders.
Our analysis of the spectrum in the MCHM showed
that the strict bound imposed by the EWPT pushes the masses of
new vectorial and fermionic 
states to values of order 4 TeV or heavier,
although this prediction can be slightly modified by varying the size of the
IR-brane kinetic terms.
A (quite) lighter KK spectrum is however expected
in those extensions of the minimal model (as above) 
in which the constraint from $Z\to b\bar b$ is weakened.
The issue of whether and how these new states can be produced at the LHC
deserves a detailed analysis, though a few promising channels of discovery
were already proposed in Ref.~\cite{Agashe:2004rs}.
There are also signals from indirect effects of the new states, such as
those in flavor physics studied by Ref.~\cite{Agashe:2004cp}.
Similarly to $\delta g_{ L b }$, there are shifts $\propto\eps^2$ in the couplings
of $t_R$ and the Higgs to $Z/W$, and between the top quark
and the Higgs or longitudinal $W/Z$. Since $t_R$
and the Higgs are (highly) composites, these effects can be as large as $\sim 10 \%$ and 
the LHC and the ILC should be able to probe them.

The analysis presented in this work for the particular case of the MCHM
hopefully clarifies some qualitative and quantitative aspects of a more
general class of composite Higgs models. 
The success of the minimal module of Ref.~\cite{Agashe:2004rs} in describing
many diverse features of electroweak and flavour physics certainly motivates
us to further investigate the idea of the Higgs as a composite~PGB.

\section*{Acknowledgments}

We are indebted to Alessandro Strumia for providing us with the fit
to the epsilon parameters used in this work and for many important discussions.
We are grateful to Martin Gr\"unewald for 
providing us with the LEP EWWG results on the epsilon parameters and for very
useful correspondence.
We thank David E. Kaplan, Guido Altarelli, Giacomo Cacciapaglia, Frank Petriello,
Alex Pomarol, Riccardo Rattazzi and Carlos Wagner for interesting discussions.
We especially thank Raman Sundrum, who has been
an extraordinary source of insight and inspiration for our work.
We would like also to thank the Aspen Center for physics for hospitality
during part of this work.
R.C. is supported by NSF grant P420-D36-2051.
K.A. was supported in part by DOE grant DE-FG02-90ER40542.


\appendix
\section{Computing the 3-point form factors}
\label{appendix:ffactors}

\begin{figure}[t]
\begin{center}
\hspace*{-1.7cm}
 \begin{minipage}[c]{1.17\linewidth}
  \vspace*{-4cm}
  \centering \epsfig{file=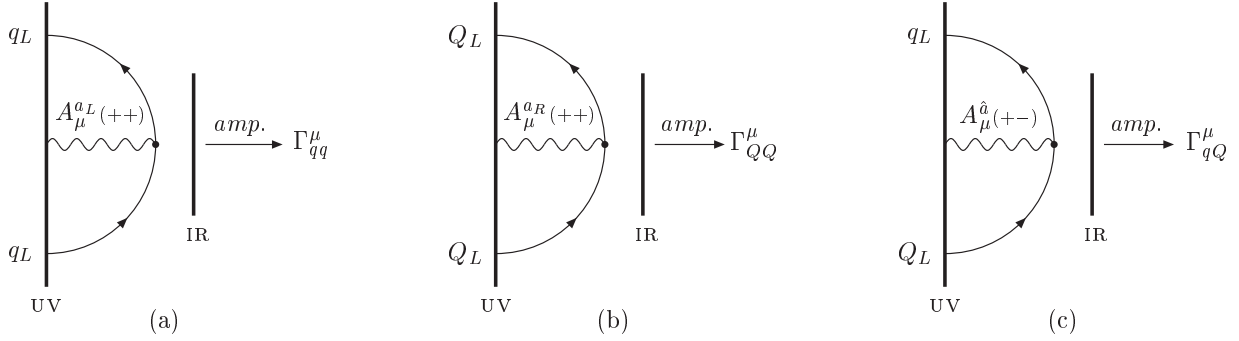,width=\linewidth}
  \vspace*{-17cm}
 \end{minipage}
\end{center}
\caption{\it 5D 3-point Green functions that define $\Gamma^\mu_{qq}$ (a), $\Gamma^\mu_{QQ}$ (b),
and $\Gamma^\mu_{qQ}$ (c).} 
\label{fig:5Ddiagrams}
\end{figure}

In this section we explain the details of the computation of the 3-point form factors
used in the text, and give their explicit expressions in terms of 5D propagators.
We use the powerful holographic technique introduced in Ref.~\cite{Agashe:2004rs},
by matching the 4D effective Lagrangian (\ref{efflag3}) to the 5D theory on the
SO(4)-invariant vacuum: $\Sigma = \Sigma_0$ (i.e. $h=0$).
We start by considering the interaction terms in eq.(\ref{efflag3})
between $\Psi_q$ and the SO(4) (unbroken) vectors
$A_\mu = A_\mu^{a_{L,R}} T^{a_{L,R}}$:
\begin{equation}
{\cal L}_{eff}^{(3)}(h=0) \supset
 \bar\Psi_q A_\mu^{a_L} T^{a_L} \left( \Gamma^{\mu ,\, q}_0 + \Gamma^{\mu ,\, q}_1 \right) \Psi_q
+\bar\Psi_q A_\mu^{a_R} T^{a_R} \left( \Gamma^{\mu ,\, q}_0 - \Gamma^{\mu ,\, q}_1 \right) \Psi_q\, .
\end{equation}
According to the holographic description, the 3-point 1PI Green functions 
$(\Gamma^{\mu ,\, q}_0 \pm \Gamma^{\mu ,\, q}_1)$ of the 4D effective Lagrangian 
correspond to the 5D 3-point functions of Fig.~\ref{fig:5Ddiagrams}(a) and (b), 
amputated of their external legs (where amputating means dividing by UV brane-UV brane propagators):
\begin{equation} \label{gq01}
\Gamma_{qq}^\mu = (\Gamma^{\mu ,\, q}_0 + \Gamma^{\mu ,\, q}_1)\, , \qquad
\Gamma_{QQ}^\mu = (\Gamma^{\mu ,\, q}_0 - \Gamma^{\mu ,\, q}_1)\, .
\end{equation}
Using the above equation, we can then extract $\Gamma^{\mu ,\, q}_{0,1}$
by computing the 5D diagrams of Fig.~\ref{fig:5Ddiagrams}.
This shows the power of the holographic description: 
the resummation of all orders in the $A_5$ insertions
can be achieved by simply calculating diagrams with no $A_5$.

A technical complication consists in deriving the various 5D fermion
propagators in the presence of the IR-brane mixing terms (\ref{massmixing}).
Resumming all these IR-brane mass insertions, 
we obtain ($q_\mu=(p-p')_\mu$): 
\begin{equation} \label{firstlist}
\Gamma_{\chi \chi}^\mu (p,p') = \int^{L_1}_{L_0}\!\! dz\, (k z)
 \left[ v^\mu_{\scriptscriptstyle (++)}(z,q)\, A_{\chi\chi}(z,p,p') + 
        \frac{1}{\ppslash} v^\mu_{\scriptscriptstyle (++)}(z,q) \frac{1}{\pslash}\, 
       B_{\chi\chi}(z,p,p') \right]\, ,
\end{equation}
\begin{equation}
\begin{split}
A_{\chi\chi}(z,p,p') =& 
 \frac{k}{G^{(+,\eta_\chi)}_{R\, \chi}(p',L_0,L_0)}
 \left( a_1^\chi(z,p') b_1^\chi(z,p) + c_1^\chi(z,p') d_1^\chi(z,p) \right)
 \frac{k}{G^{(+,\eta_\chi)}_{R\, \chi}(p,L_0,L_0)}\, , \\
B_{\chi\chi}(z,p,p') =& 
 \frac{k}{G^{(+,\eta_\chi)}_{R\, \chi}(p',L_0,L_0)}
 \left( a_2^\chi(z,p') b_2^\chi(z,p) + c_2^\chi(z,p') d_2^\chi(z,p) \right)
 \frac{k}{G^{(+,\eta_\chi)}_{R\, \chi}(p,L_0,L_0)}\, , 
\end{split}
\end{equation}
where $\chi = q, Q$, $\eta_q = +$, $\eta_Q = -$, and
\begin{align}
a_1^q (z,p) =& G^{(++)}_{R\, q}(p,L_0,z) +
 \frac{\left(pL_1 \right)^2 \m^\dagger\m \,
       G^{(++)}_{R\, q}(p,L_0,L_1) G^{(-+)}_{L\, q^u}(p,L_1,L_1) G^{(-+)}_{R\, q}(p,L_1,z) }%
      {1-\left(pL_1 \right)^2 \m^\dagger\m \, 
       G^{(-+)}_{R\, q}(p,L_1,L_1) G^{(-+)}_{L\, q^u}(p,L_1,L_1) }\, , \\
a_2^q (z,p) =& \tilde G^{(--)}_{L\, q}(p,L_0,z) +
 \frac{\left(pL_1 \right)^2 \m^\dagger\m \,
       G^{(++)}_{R\, q}(p,L_0,L_1) G^{(-+)}_{L\, q^u}(p,L_1,L_1) \tilde G^{(+-)}_{L\, q}(p,L_1,z) }%
      {1- \left(pL_1 \right)^2 \m^\dagger\m \, 
       G^{(-+)}_{R\, q}(p,L_1,L_1) G^{(-+)}_{L\, q^u}(p,L_1,L_1) }\, ,\\
b_1^q (z,p) =& G^{(++)}_{R\, q}(p,z,L_0) +
 \frac{\left(pL_1 \right)^2 \m^\dagger\m \,
       G^{(-+)}_{R\, q}(p,z,L_1) G^{(-+)}_{L\, q^u}(p,L_1,L_1) G^{(++)}_{R\, q}(p,L_1,L_0) }%
      {1-\left(pL_1 \right)^2 \m^\dagger\m \, 
       G^{(-+)}_{R\, q}(p,L_1,L_1) G^{(-+)}_{L\, q^u}(p,L_1,L_1) }\, , \\
b_2^q (z,p) =& \tilde G^{(++)}_{R\, q}(p,z,L_0) +
 \frac{\left(pL_1 \right)^2 \m^\dagger\m \,
       \tilde G^{(-+)}_{R\, q}(p,z,L_1) G^{(-+)}_{L\, q^u}(p,L_1,L_1) G^{(++)}_{R\, q}(p,L_1,L_0) }%
      {1-\left(pL_1 \right)^2 \m^\dagger\m \,
       G^{(-+)}_{R\, q}(p,L_1,L_1) G^{(-+)}_{L\, q^u}(p,L_1,L_1) }\, , \\[0.3cm]
c_1^q (z,p) =& 
 \frac{ \m L_1 \, G^{(++)}_{R\, q}(p,L_0,L_1) \tilde G^{(+-)}_{R\, q^u}(p,L_1,z) }%
      {1-\left(pL_1 \right)^2 \m^\dagger\m \, 
       G^{(-+)}_{R\, q}(p,L_1,L_1) G^{(-+)}_{L\, q^u}(p,L_1,L_1) }\, , \\
c_2^q (z,p) =&
 \frac{ \m L_1 \, p^2 G^{(++)}_{R\, q}(p,L_0,L_1) G^{(-+)}_{L\, q^u}(p,L_1,z) }%
      {1-\left(pL_1 \right)^2 \m^\dagger\m \,
       G^{(-+)}_{R\, q}(p,L_1,L_1) G^{(-+)}_{L\, q^u}(p,L_1,L_1) }\, , \\
d_1^q (z,p) =& 
 \frac{ \m^\dagger L_1 \, \tilde G^{(-+)}_{L\, q^u}(p,z,L_1) G^{(++)}_{R\, q}(p,L_1,L_0) }%
      {1-\left(pL_1 \right)^2 \m^\dagger\m \,
       G^{(-+)}_{R\, q}(p,L_1,L_1) G^{(-+)}_{L\, q^u}(p,L_1,L_1) }\, , \\
d_2^q (z,p) =& 
 \frac{\m^\dagger L_1 \, p^2  G^{(-+)}_{L\, q^u}(p,z,L_1) G^{(++)}_{R\, q}(p,L_1,L_0) }%
      {1-\left(pL_1 \right)^2 \m^\dagger\m \, 
       G^{(-+)}_{R\, q}(p,L_1,L_1) G^{(-+)}_{L\, q^u}(p,L_1,L_1) }\, , 
\end{align}   
\begin{align}
a_1^Q (z,p) =& G^{(+-)}_{R\, Q}(p,L_0,z) +
 \frac{ L_1^2 \, \M^\dagger\M \,
   \tilde G^{(-+)}_{L\, Q}(p,L_0,L_1) G^{(++)}_{R\, Q^u}(p,L_1,L_1) \tilde G^{(--)}_{R\, Q}(p,L_1,z) }%
      {1-\left(pL_1 \right)^2 \M^\dagger\M \,
       G^{(++)}_{L\, Q}(p,L_1,L_1) G^{(++)}_{R\, Q^u}(p,L_1,L_1) }\, , \\
a_2^Q (z,p) =& \tilde G^{(-+)}_{L\, Q}(p,L_0,z) +
 \frac{ \left(pL_1\right)^2 \M^\dagger\M \,
   \tilde G^{(-+)}_{L\, Q}(p,L_0,L_1) G^{(++)}_{R\, Q^u}(p,L_1,L_1) G^{(++)}_{L\, Q}(p,L_1,z) }%
      {1-\left(pL_1 \right)^2 \M^\dagger\M \,
       G^{(++)}_{L\, Q}(p,L_1,L_1) G^{(++)}_{R\, Q^u}(p,L_1,L_1) }\, , \\
b_1^Q (z,p) =& G^{(+-)}_{R\, Q}(p,z,L_0) +
 \frac{ L_1^2 \, \M^\dagger\M \,
   \tilde G^{(++)}_{L\, Q}(p,z,L_1) G^{(++)}_{R\, Q^u}(p,L_1,L_1) \tilde G^{(+-)}_{R\, Q}(p,L_1,L_0) }%
      {1-\left(pL_1 \right)^2 \M^\dagger\M \,
       G^{(++)}_{L\, Q}(p,L_1,L_1) G^{(++)}_{R\, Q^u}(p,L_1,L_1) }\, , \\
b_2^Q (z,p) =& \tilde G^{(+-)}_{R\, Q}(p,z,L_0) +
 \frac{ \left(pL_1\right)^2 \M^\dagger\M \,
    G^{(++)}_{L\, Q}(p,z,L_1) G^{(++)}_{R\, Q^u}(p,L_1,L_1) \tilde G^{(+-)}_{R\, Q}(p,L_1,L_0) }%
 {1-\left(pL_1 \right)^2 \M^\dagger\M G^{(++)}_{L\, Q}(p,L_1,L_1) G^{(++)}_{R\, Q^u}(p,L_1,L_1) }\, , \\[0.3cm]
c_1^Q (z,p) =& \frac{ \M L_1 \tilde G^{(-+)}_{L\, Q}(p,L_0,L_1) G^{(++)}_{R\, Q^u}(p,L_1,z)  }%
                    {1-\left(pL_1 \right)^2 \M^\dagger\M \, 
                     G^{(++)}_{L\, Q}(p,L_1,L_1) G^{(++)}_{R\, Q^u}(p,L_1,L_1) }\, , \\
c_2^Q (z,p) =& \frac{ \M L_1 \tilde G^{(-+)}_{L\, Q}(p,L_0,L_1) \tilde G^{(--)}_{L\, Q^u}(p,L_1,z)  }%
                    {1-\left(pL_1 \right)^2 \M^\dagger\M \,
                     G^{(++)}_{L\, Q}(p,L_1,L_1) G^{(++)}_{R\, Q^u}(p,L_1,L_1) }\, , \\
d_1^Q (z,p) =& \frac{ \M^\dagger L_1 G^{(++)}_{R\, Q^u}(p,z,L_1) \tilde G^{(+-)}_{R\, Q}(p,L_1,L_0)  }%
                    {1-\left(pL_1 \right)^2 \M^\dagger\M \, 
                     G^{(++)}_{L\, Q}(p,L_1,L_1) G^{(++)}_{R\, Q^u}(p,L_1,L_1) }\, , \\
\label{lastlist}
d_2^Q (z,p) =& \frac{ \M^\dagger L_1 \tilde 
                     G^{(++)}_{R\, Q^u}(p,z,L_1) \tilde G^{(+-)}_{R\, Q}(p,L_1,L_0)  }%
                    {1-\left(pL_1 \right)^2 \M^\dagger\M \, 
                     G^{(++)}_{L\, Q}(p,L_1,L_1) G^{(++)}_{R\, Q^u}(p,L_1,L_1) }\, .
\end{align}
Here $G_{L,R}$ are defined as the left- and right-handed
components of the 5D propagator $S(p,z,z')$ of a bulk fermion with mass $c k$
between two points $z$, $z'$ along the fifth dimension 
(see for example~\cite{Gherghetta:2000kr,Contino:2003ve}):
\begin{equation}
S(p,z,z')= (k^2 z z')^{5/2}
 \left[ \pslash +\gamma^5 \left(\partial_z +\frac{1}{2z}\right)+\frac{c}{z} \right]
 \left[ P_R\, G_R(p,z,z')+P_L\, G_L(p,z,z') \right]\, ,
\end{equation}
where $P_{R,L}=(1\pm\gamma^5)/2$.
We have also defined
\begin{equation}
\begin{split}
\tilde G_{L}(z,z') &= \left[ - \partial_z + \frac{(c - 1/2)}{z} \right] G_{L}(z,z')\, , \\
\tilde G_{R}(z,z') &= \left[ + \partial_z + \frac{(c + 1/2)}{z} \right] G_{R}(z,z')\, .
\end{split}
\end{equation}
The 4-vector $v_{\scriptscriptstyle (+\pm)}^\mu(z,q)$ is defined as
\begin{equation} \label{vmu}
v^\mu_{\scriptscriptstyle (+\pm)}(z,q) = 
  \left[ \left(P_L\right)^\mu_\rho + \frac{G_T^{(+\pm)}(q,L_0,z)}{G_T^{(+\pm)}(q,L_0,L_0)} 
         \left(P_T\right)^\mu_\rho \right] \gamma^\rho \, ,
\end{equation}
and satisfies
\begin{equation}
q_\mu\, v_{\scriptscriptstyle (+\pm)}^\mu(z,q) = \qslash \, , \qquad
 v_{\scriptscriptstyle (+\pm)}^\mu(z,q^2=0) = \gamma^\mu \, .
\end{equation}
In eq.(\ref{vmu}) $G_T$ stands for the transverse component of the 5D gauge propagator
\begin{equation}
G^\mu_\nu (q,z,z') = 
 \left(P_L\right)^\mu_\nu G_L(q,z,z') + \left(P_T\right)^\mu_\nu G_T(q,z,z')\, ,
\end{equation}
and $(P_T)_{\mu\nu} = \eta_{\mu\nu}-q_\mu q_\nu/q^2$, $(P_L)_{\mu\nu} =q_\mu q_\nu/q^2$
are respectively the transverse and longitudinal projectors.

Using eq.(\ref{gq01}), together with eqs.(\ref{firstlist})-(\ref{lastlist}), 
one can compute the form factors
$\Gamma^{\mu ,\, q}_{0,1}$. In order to extract $\Gamma^{\mu ,\, u}_{0,1}$
we notice that they are related to the previous ones by exchanging
$c_q \leftrightarrow c_u$ and $L \leftrightarrow R$, that is:
\begin{equation}
\Gamma^{\mu ,\, u}_{0,1}(p,p') = \Gamma^{\mu ,\, q}_{0,1}(p,p') 
 \left(c_q \leftrightarrow -c_u \right)\, .
\end{equation}

Finally, we notice that the Ward identity (\ref{WI}) leads to a useful integral
representation of the form factors $\Pi_{0,1}$:
\begin{equation} \label{intrep}
\begin{split}
\Pi^q_0(p)+ \Pi^q_1(p) =& \int^{L_1}_{L_0} \!\! dz\, (kz) 
 \left( A_{qq}(z,p,p')+\frac{1}{p^2} B_{qq}(z,p,p') \right)\, , \\
\Pi^q_0(p)- \Pi^q_1(p) =& \int^{L_1}_{L_0} \!\! dz\, (kz) 
 \left( A_{QQ}(z,p,p')+\frac{1}{p^2} B_{QQ}(z,p,p') \right)\, ,
\end{split}
\end{equation}
and similarly for $\Pi_{0,1}^u$. 
For specific values of $c_{q,u}$, we checked that the expression
of the $\Pi$'s as obtained from eqs.(\ref{intrep})
coincide with that derived in~\cite{Agashe:2004rs} 
from the two-point functions.

We now turn to the computation of the other form factor relevant to the computation
of $\delta g_{Lb}$ and $\Delta\rho$: $\Gamma^{\mu}_2$.
To this end we consider the interaction terms in eq.(\ref{efflag3}) among $q_L$, $Q_L$
and the SO(5)/SO(4) (broken) vectors $A_\mu = A_\mu^{\hat a} T^{\hat a}$,
again setting $\Sigma = \Sigma_0$ ($h=0$):
\begin{equation}
{\cal L}_{eff}^{(3)}(h=0) \supset \bar q_L A_\mu^{\hat a} \big(T^{\hat a}\big)_{qQ}
 \left[ \Gamma^{\mu ,\, q}_0 + \Gamma^{\mu ,\, q}_2
        -2 i \,\Gamma^{\mu ,\, q}_3 \right] Q_L + \text{h.c.}
\end{equation}
Holography prescribes that the 1PI Green function 
$[\Gamma^{\mu ,\, q}_0 + \Gamma^{\mu ,\, q}_2-2 i \,\Gamma^{\mu ,\, q}_3 ]$
corresponds to the 5D 3-point function of Fig.~\ref{fig:5Ddiagrams}(c) amputated of its external legs
(as before, amputating means dividing  by the corresponding UV brane-UV brane propagators).
To get rid of $\Gamma^{\mu ,\, q}_3$ we simply extract the Hermitian part of the diagram
of Fig.~\ref{fig:5Ddiagrams}(c), since all form factors $\Gamma^\mu_i$ are Hermitian 
(see eq.(\ref{hermiticity})):
\begin{equation} \label{gq23}
\Real\{\Gamma^\mu_{qQ}\} = \Gamma^{\mu ,\, q}_0 + \Gamma^{\mu ,\, q}_2\, .
\end{equation}
We find:
\begin{equation}
\Real\{\Gamma_{qQ}^\mu (p,p')\} = \int^{L_1}_{L_0}\!\! dz\, (k z)
 \left[ v^\mu_{\scriptscriptstyle (+-)}(z,q)\, A_{qQ}(z,p,p') + 
        \frac{1}{\ppslash} v^\mu_{\scriptscriptstyle (+-)}(z,q) \frac{1}{\pslash}\, 
       B_{qQ}(z,p,p') \right]\, ,
\end{equation}
\begin{equation} \label{ABqQ}
\begin{split}
A_{qQ}(z,p,p') =& 
 \frac{k}{G^{(++)}_{R\, q}(p',L_0,L_0)}
 \left( a_1^q(z,p') b_1^Q(z,p) + \Real\{c_1^q(z,p') d_1^Q(z,p)\} \right)
 \frac{k}{G^{(+-)}_{R\, Q}(p,L_0,L_0)}\, , \\
B_{qQ}(z,p,p') =& 
 \frac{k}{G^{(++)}_{R\, q}(p',L_0,L_0)}
 \left( a_2^q(z,p') b_2^Q(z,p) + \Real\{c_2^q(z,p') d_2^Q(z,p)\} \right)
 \frac{k}{G^{(+-)}_{R\, Q}(p,L_0,L_0)}\, .
\end{split}
\end{equation}
Using eqs.(\ref{gq23})-(\ref{ABqQ}) and the expression of $\Gamma^{\mu ,\, q}_0$ found
previously, one can then deduce $\Gamma^{\mu ,\, q}_2$.
The form factor $\Gamma^{\mu ,\, u}_2$ can be
obtained by exchanging $c_q \leftrightarrow c_u$ and $L \leftrightarrow R$:
\begin{equation}
\Gamma^{\mu ,\, u}_2(p,p') = \Gamma^{\mu ,\, q}_2(p,p') (c_q\leftrightarrow -c_u)\, .
\end{equation}

From the expression for $\Gamma^{\mu}_2$
computed above, and using eq.(\ref{G23})
we can finally write down the explicit result for 
$\Gamma_{2}(p)$, $\Delta_{2}(p)$:
\begin{align}
\begin{split}
\Gamma^{q}_{2}(p) =& \int^{L_1}_{L_0}\!\! dz\, (kz)
 \bigg\{ A_{qQ}(z,p,p)+\frac{1}{p^2} B_{qQ}(z,p,p) \\
  &-\frac{1}{2} \left[ A_{qq}(z,p,p)+\frac{1}{p^2} B_{qq}(z,p,p) +
                A_{QQ}(z,p,p)+\frac{1}{p^2} B_{QQ}(z,p,p) \right] \bigg\}\, ,
\end{split} \\
\Delta^q_{2}(p) =& \int^{L_1}_{L_0}\!\! dz\, (kz) \frac{1}{p^4}
 \left\{ B_{qQ}(z,p,p) - \frac{1}{2} \left( B_{qq}(z,p,p)+B_{QQ}(z,p,p)\right)\right\}\, , \\[0.2cm]
\Gamma^{u}_{2}(p) =& \Gamma^{q}_{2}(p) \left(c_q\leftrightarrow -c_u \right)\, , \\
\Delta^{u}_{2}(p) =& \Delta^{q}_{2}(p) \left(c_q\leftrightarrow -c_u \right)\, .
\end{align}
%



\begin{thebibliography}{99}

\bibitem{GK}
D.~B.~Kaplan and H.~Georgi,
Phys.\ Lett.\ B {\bf 136}, 183 (1984);
 B {\bf 136}, 187 (1984); \\
H.~Georgi, D.~B.~Kaplan and P.~Galison,
Phys.\ Lett.\ B {\bf 143}, 152 (1984); \\
H.~Georgi and D.~B.~Kaplan,
Phys.\ Lett.\ B {\bf 145}, 216 (1984); \\
M.~J.~Dugan, H.~Georgi and D.~B.~Kaplan,
Nucl.\ Phys.\ B {\bf 254}, 299 (1985).

\bibitem{TC}
S.~Weinberg,
Phys.\ Rev.\ D {\bf 13}, 974 (1976);
Phys.\ Rev.\ D {\bf 19}, 1277 (1979);
L.~Susskind,
Phys.\ Rev.\ D {\bf 20}, 2619 (1979).

\bibitem{Contino:2003ve}
R.~Contino, Y.~Nomura and A.~Pomarol,
Nucl.\ Phys.\ B {\bf 671}, 148 (2003)
[arXiv:hep-ph/0306259].

\bibitem{Agashe:2004rs}
K.~Agashe, R.~Contino and A.~Pomarol,
Nucl.\ Phys.\ B {\bf 719}, 165 (2005)
[arXiv:hep-ph/0412089].

\bibitem{Kaplan:1991dc}
D.~B.~Kaplan,
Nucl.\ Phys.\ B {\bf 365}, 259 (1991).

\bibitem{Grossman:1999ra}
Y.~Grossman and M.~Neubert,
Phys.\ Lett.\ B {\bf 474}, 361 (2000)
[arXiv:hep-ph/9912408].

\bibitem{Gherghetta:2000qt}
T.~Gherghetta and A.~Pomarol,
Nucl.\ Phys.\ B {\bf 586}, 141 (2000)
[arXiv:hep-ph/0003129].

\bibitem{Huber:2000ie}
S.~J.~Huber and Q.~Shafi,
Phys.\ Lett.\ B {\bf 498}, 256 (2001)
[arXiv:hep-ph/0010195];
S.~J.~Huber,
Nucl.\ Phys.\ B {\bf 666}, 269 (2003)
[arXiv:hep-ph/0303183].

\bibitem{Agashe:2003zs}
K.~Agashe, A.~Delgado, M.~J.~May and R.~Sundrum,
JHEP {\bf 0308}, 050 (2003)
[arXiv:hep-ph/0308036].

\bibitem{Agashe:2004cp}
K.~Agashe, G.~Perez and A.~Soni,
Phys.\ Rev.\ Lett.\  {\bf 93}, 201804 (2004)
[arXiv:hep-ph/0406101];
arXiv:hep-ph/0408134.

\bibitem{Dimopoulos:1979es}
S.~Dimopoulos and L.~Susskind,
Nucl.\ Phys.\ B {\bf 155}, 237 (1979).

\bibitem{gaugeHiggs}
D.~B.~Fairlie,
Phys.\ Lett.\ B {\bf 82}, 97 (1979);
N.~S.~Manton,
Nucl.\ Phys.\ B {\bf 158}, 141 (1979);
D.~B.~Fairlie,
J.\ Phys.\ G {\bf 5}, L55 (1979);
P.~Forgacs and N.~S.~Manton,
Commun.\ Math.\ Phys.\  {\bf 72}, 15 (1980);
S.~Randjbar-Daemi, A.~Salam and J.~A.~Strathdee,
Nucl.\ Phys.\ B {\bf 214}, 491 (1983).

\bibitem{Randall:1999ee}
L.~Randall and R.~Sundrum,
Phys.\ Rev.\ Lett.\  {\bf 83}, 3370 (1999)
[arXiv:hep-ph/9905221].

\bibitem{Agashe:2005vg}
K.~Agashe, R.~Contino and R.~Sundrum,
arXiv:hep-ph/0502222.

\bibitem{Scrucca:2003ra}
C.~A.~Scrucca, M.~Serone and L.~Silvestrini,
Nucl.\ Phys.\ B {\bf 669}, 128 (2003)
[arXiv:hep-ph/0304220].

\bibitem{Dobrescu:1999gv}
B.~A.~Dobrescu,
Phys.\ Rev.\ D {\bf 63}, 015004 (2001)
[arXiv:hep-ph/9908391].

\bibitem{PT}
M.~E.~Peskin and T.~Takeuchi,
Phys.\ Rev.\ Lett.\  {\bf 65}, 964 (1990);
Phys.\ Rev.\ D {\bf 46}, 381 (1992).

\bibitem{Barbieri:1999tm}
R.~Barbieri and A.~Strumia,
Phys.\ Lett.\ B {\bf 462}, 144 (1999)
[arXiv:hep-ph/9905281].

\bibitem{Barbieri:2000gf}
R.~Barbieri and A.~Strumia,
arXiv:hep-ph/0007265.

\bibitem{Katz:2005au}
E.~Katz, A.~E.~Nelson and D.~G.~E.~Walker,
JHEP {\bf 0508}, 074 (2005)
[arXiv:hep-ph/0504252].

\bibitem{Chacko:2005pe}
Z.~Chacko, H.~S.~Goh and R.~Harnik,
arXiv:hep-ph/0506256.

\bibitem{Georgi:1975tz}
H.~Georgi and A.~Pais,
Phys.\ Rev.\ D {\bf 12}, 508 (1975).

\bibitem{Arkani-Hamed:2001nc}
N.~Arkani-Hamed, A.~G.~Cohen and H.~Georgi,
Phys.\ Lett.\ B {\bf 513}, 232 (2001)
[arXiv:hep-ph/0105239];

\bibitem{LH}
N.~Arkani-Hamed, A.~G.~Cohen, E.~Katz, A.~E.~Nelson, T.~Gregoire and J.~G.~Wacker,
JHEP {\bf 0208}, 021 (2002)
[arXiv:hep-ph/0206020];
N.~Arkani-Hamed, A.~G.~Cohen, E.~Katz and A.~E.~Nelson,
JHEP {\bf 0207}, 034 (2002)
[arXiv:hep-ph/0206021];
D.~E.~Kaplan and M.~Schmaltz,
JHEP {\bf 0310}, 039 (2003)
[arXiv:hep-ph/0302049].
For a review see:
M.~Schmaltz and D.~Tucker-Smith,
arXiv:hep-ph/0502182,
and references therein.

\bibitem{Csaki:2002qg}
C.~Csaki, J.~Hubisz, G.~D.~Kribs, P.~Meade and J.~Terning,
Phys.\ Rev.\ D {\bf 67}, 115002 (2003)
[arXiv:hep-ph/0211124].

\bibitem{Hewett:2002px}
J.~L.~Hewett, F.~J.~Petriello and T.~G.~Rizzo,
JHEP {\bf 0310}, 062 (2003)
[arXiv:hep-ph/0211218].

\bibitem{recentLH}For a recent analysis, see
Z.~Han and W.~Skiba,
Phys.\ Rev.\ D {\bf 72}, 035005 (2005)
[arXiv:hep-ph/0506206];
G.~Marandella, C.~Schappacher and A.~Strumia,
Phys.\ Rev.\ D {\bf 72}, 035014 (2005)
[arXiv:hep-ph/0502096].
For more references, see the review by
M.~Schmaltz and D.~Tucker-Smith in~\cite{LH}.

\bibitem{Cheng:2003ju}
H.~C.~Cheng and I.~Low,
JHEP {\bf 0309}, 051 (2003)
[arXiv:hep-ph/0308199];
JHEP {\bf 0408}, 061 (2004)
[arXiv:hep-ph/0405243].

\bibitem{Thaler:2005en}
J.~Thaler and I.~Yavin,
JHEP {\bf 0508}, 022 (2005)
[arXiv:hep-ph/0501036].

\bibitem{Katz:2003sn}
E.~Katz, J.~y.~Lee, A.~E.~Nelson and D.~G.~E.~Walker,
arXiv:hep-ph/0312287.

\bibitem{Batra:2004ah}
P.~Batra and D.~E.~Kaplan,
JHEP {\bf 0503}, 028 (2005)
[arXiv:hep-ph/0412267].

\bibitem{hosotani}
Y.~Hosotani,
Phys.\ Lett.\ B {\bf 126}, 309 (1983);
Phys.\ Lett.\ B {\bf 129}, 193 (1983).

\bibitem{Barbieri:2003pr}
R.~Barbieri, A.~Pomarol and R.~Rattazzi,
Phys.\ Lett.\ B {\bf 591}, 141 (2004)
[arXiv:hep-ph/0310285].

\bibitem{Contino:2004vy}
R.~Contino and A.~Pomarol,
JHEP {\bf 0411}, 058 (2004)
[arXiv:hep-th/0406257].

\bibitem{Maldacena:1997re}
J.~M.~Maldacena,
Adv.\ Theor.\ Math.\ Phys.\  {\bf 2}, 231 (1998)
[Int.\ J.\ Theor.\ Phys.\  {\bf 38}, 1113 (1999)]
[arXiv:hep-th/9711200];
S.~S.~Gubser, I.~R.~Klebanov and A.~M.~Polyakov,
Phys.\ Lett.\ B {\bf 428}, 105 (1998)
[arXiv:hep-th/9802109];
E.~Witten,
Adv.\ Theor.\ Math.\ Phys.\  {\bf 2}, 253 (1998)
[arXiv:hep-th/9802150].

\bibitem{LEPEWWG}
The LEP Electroweak Working Group, CERN-PH-EP/2004-069 and arXiv:hep-ex/0412015
(December 2004), updated for 2005 winter conferences.

\bibitem{eps123}
G.~Altarelli and R.~Barbieri,
Phys.\ Lett.\ B {\bf 253}, 161 (1991);
G.~Altarelli, R.~Barbieri and S.~Jadach,
Nucl.\ Phys.\ B {\bf 369}, 3 (1992)
[Erratum-ibid.\ B {\bf 376}, 444 (1992)].
  
\bibitem{epsb}
G.~Altarelli, R.~Barbieri and F.~Caravaglios,
Nucl.\ Phys.\ B {\bf 405}, 3 (1993).

\bibitem{TopaZ0}
G.~Montagna, F.~Piccinini, O.~Nicrosini, G.~Passarino and R.~Pittau,
Nucl.\ Phys.\ B {\bf 401}, 3 (1993);
Comput.\ Phys.\ Commun.\  {\bf 76}, 328 (1993);
G.~Montagna, O.~Nicrosini, G.~Passarino and F.~Piccinini,
Comput.\ Phys.\ Commun.\  {\bf 93}, 120 (1996)
[arXiv:hep-ph/9506329];
Comput.\ Phys.\ Commun.\  {\bf 117}, 278 (1999)
[arXiv:hep-ph/9804211].

\bibitem{Barbieri:2004qk}
R.~Barbieri, A.~Pomarol, R.~Rattazzi and A.~Strumia,
Nucl.\ Phys.\ B {\bf 703}, 127 (2004)
[arXiv:hep-ph/0405040].

\bibitem{Ale}
Alessandro Strumia, private communication.
The data used in the fit are those of LEP1 
(see Table 2 of Ref.~\cite{Barbieri:2004qk}),
and those from Atomic Parity Violation (APV) 
(see Ref.~\cite{Barbieri:2004qk}, Table 3).
The NuTeV data have not been included,
and in any case their inclusion would not significantly 
change the results of the fit.

\bibitem{Group:2005cc}
t.~T.~E.~Group  [the D0 Collaboration],
arXiv:hep-ex/0507091.

\bibitem{Choudhury:2001hs}
D.~Choudhury, T.~M.~P.~Tait and C.~E.~M.~Wagner,
Phys.\ Rev.\ D {\bf 65}, 053002 (2002)
[arXiv:hep-ph/0109097].

\bibitem{Gherghetta:2000kr}
T.~Gherghetta and A.~Pomarol,
Nucl.\ Phys.\ B {\bf 602}, 3 (2001)
[arXiv:hep-ph/0012378].


\end{thebibliography}
\end{document}